\documentclass[showpacs,amssymb,10pt,reprint,aps,prd,longbibliography,nofootinbib,superscriptaddress,floatfix,nobibnotes]{revtex4-1}

\usepackage{graphicx,epsfig,amssymb} 
\usepackage{amsmath,amsfonts, times}
\usepackage{bm} 

\usepackage{epstopdf}
\usepackage[linktocpage,colorlinks]{hyperref}
\usepackage[caption=false]{subfig}
\usepackage[usenames]{color}     
\usepackage{natbib}
\usepackage{subfig}
\usepackage[utf8x]{inputenc}
\usepackage{enumerate}
\usepackage[normalem]{ulem}

\begin{document}
\title{\large Static boson stars in the Einstein-Friedberg-Lee-Sirlin theory and their astrophysical images}

\author{Pedro L. Brito de Sá}
\email{pedro.sa@icen.ufpa.br}
\affiliation{Programa de P\'os-Gradua\c{c}\~{a}o em F\'{\i}sica, Universidade 
	Federal do Par\'a, 66075-110, Bel\'em, Par\'a, Brazil.}

\author{Haroldo C. D. Lima Jr.}
\email{haroldo.lima@ufma.br}
    \affiliation{Programa de P\'os-Gradua\c{c}\~{a}o em F\'{\i}sica, Universidade 
	Federal do Par\'a, 66075-110, Bel\'em, Par\'a, Brazil.}
	\affiliation{Departamento de F{\'i}sica - CCET, Universidade Federal do Maranh{\~a}o, Campus Universit{\'a}rio do Bacanga, 65080-805, S{\~a}o Lu{\'i}s, Maranh{\~a}o, Brazil.}
	
\author{Carlos A. R. Herdeiro}
\email{herdeiro@ua.pt}
\affiliation{Departamento de Matem\'atica da Universidade de Aveiro and Centre for Research and Development  in Mathematics and Applications (CIDMA), Campus de Santiago, 3810-183 Aveiro, Portugal.}

\author{Lu\'is C. B. Crispino}
\email{crispino@ufpa.br}
\affiliation{Programa de P\'os-Gradua\c{c}\~{a}o em F\'{\i}sica, Universidade 
	Federal do Par\'a, 66075-110, Bel\'em, Par\'a, Brazil.}

\begin{abstract}
We investigate the static boson star solutions in the so-called Einstein-Friedberg-Lee-Sirlin (E-FLS) theory, performing a complete analysis of the solution space in this model. We study the phenomenological aspects of E-FLS stars, for instance, by investigating the timelike and null geodesics with an emphasis on the analysis of circular timelike orbits and light rings. In order to study the astrophysical signatures of such stars, their images were obtained considering them surrounded by a geometrically thin accretion disk. Our results comprise two different models of accretion disks, namely the optically thin and thick disk models. We present a selection of our findings for the astrophysical images of E-FLS stars and discuss their relevance as a possible black hole mimicker.
\end{abstract}

\date{June 2024}

\maketitle

\section{Introduction}
\label{sec:int}
The last decades have represented a golden age concerning observational data on the strong field regime for the theory of General Relativity (GR).
The detection of gravitational waves (GW) by the collaboration LIGO/VIRGO~\cite{LIGO-VIRGO2016} inaugurated the era of GW astronomy, i.e. the exploration of the cosmos through a new type of radiation other than the electromagnetic one. Up to date, several events have been observed through GWs, such as, the merger of black hole-black hole and black hole-neutron star systems~\cite{LIGO-VIRGO}.
In 2019, the Event Horizon Telescope (EHT) collaboration imaged, for the first time ever, the structure close to the central regions of both the M87 and Milky Way galaxies~\cite{M87_1:2019, M87_2:2019, M87_3:2019, M87_4:2019, M87_5:2019, M87_6:2019, sgra_1:2022, sgra_2:2022, sgra_3:2022, sgra_4:2022, sgra_5:2022, sgra_6:2022}, using Very Long Baseline Interferometry (VLBI) techniques. The images present a inner dark region which is compatible to the shadow of a supermassive compact object. 

In the near future, additional GW detectors will become operational, improving the precision of GW astronomy. Moreover the next generation Event Horizon Telescope (ng-EHT) is planed to increase the current baselines in order to capture images with a higher level of details.
The forthcoming data on the gravitational and electromagnetic channels can be used to constrain models that challenge the universality of the Kerr black hole (BH) hypothesis~\cite{Kerr_Hypothesis}. The departure from the Kerr hypothesis may arise from either (i) alternative black hole models, in particular, described by additional parameters beyond mass and angular momentum within the Kerr metric, or (ii) compact objects lacking a horizon yet exhibiting characteristics that closely resemble those of BHs.

Among such horizonless BH mimickers, non-topological solitonic solutions known as \textit{boson stars} stand as robust viable candidates. Some boson stars are linearly stable in regions of their parameter space~\cite{LeePang,Santos:2024vdm}. Additionally, they admit a robust formation mechanism dubbed as \textit{gravitational cooling}~\cite{grav_cool1,grav_cool2,grav_cool3}.
The pioneering work on boson star solutions dates back to 1968 with the studies of Kaup~\cite{Kaup}, where he obtained the boson star solutions in the context of GR in the presence of a complex scalar field with a potential without self interaction terms, known nowadays as mini-boson stars. After that, boson star solutions for different sorts of potential were studied in the literature. Additionally, analogous compact objects in GR composed of massive vector fields have been discovered more recently and they are known as Proca stars~\cite{Proca1,Proca2}.

In 1976, a pioneer work showed that solitons in flat spacetime can emerge as solutions of a system composed of a complex scalar field coupled with a real scalar field, known as Friedberg-Lee-Sirlin (FLS) model~\cite{FLS}. Remarkably, the FLS model provide a simple example of a renormalizable theory. The masses of the complex and real scalar fields are generated due to the coupling between them. The FLS model minimally coupled to the Einstein gravity, dubbed as E-FLS theory, has been studied in different works. In Ref.~\cite{Rotating_FLS}, the rotating boson stars and the Kerr BHs with synchronized scalar hair were studied in the context of FLS model. The non-rotating gauged boson stars and gauged Q-balls were obtained in Ref.~\cite{U1_FLS}. Furthermore, a vector version of the FLS model was considered in Ref~\cite{Proca-Higgs}, motivated by the need to avoid hyperbolicity issues commonly encountered in the standard Proca models.

Although the rotating and gauged solutions in the E-FLS theory have been studied, a detailed analysis of the space of solutions in the non-rotating and ungauged case has not been presented in the literature, as far as we are aware. Additionally, the study of the phenomenology of such static and ungauged E-FLS stars is of uttermost importance and has not been done so far. In this paper we provide a complete analysis of the solution space in the static case, exploring the full range of parameter within the E-FLS model. Moreover, we also study the timelike and lightlike geodesics around E-FLS stars. Finally, we obtain the astrophysical images of such E-FLS stars when surrounded by different accretion disk models, which are important in the light of the recent EHT results. 

This paper is organized as follows. In Sec.~\ref{sec:model}, we present the FLS model minimally coupled to gravity, the field equations and the appropriate boundary conditions. In Sec.~\ref{numerical_sol}, we discuss the numerical method and present a selection of our numerical results for boson star solutions. In Sec.~\ref{Geo_ana} we analyze the timelike and lightlike geodesics around such stars, obtaining, for instance, the circular timelike orbits and the light rings. In Sec.~\ref{Shadows}, we apply the backwards ray-tracing method in order to obtain the astrophysical images of these stars, when surrounded by an accretion disk. We present our final remarks in Sec.~\ref{Conclusions}. In the remainder of this paper, we adopt geometrical units such that $c=\hbar=1$ and the metric signature is $+2$. 

\section{The E-FLS theory}
\label{sec:model}
We explore solitonic solutions in (3+1)-dimensions with two scalar fields (a complex field and a real field), minimally coupled to gravity, whose interaction is governed by the FLS model~\cite{FLS}. The E-FLS theory studied here is defined by the following action:
\begin{align}
\label{action}S=\int d^4x\sqrt{-g}\left(\frac{R}{2\kappa^2}-\mathcal{L}_{m}\right), \qquad \kappa^2\equiv 8\pi G,
\end{align}
where $R$ is the Ricci scalar, $g$ is the metric determinant, and 
\begin{align}
&\mathcal{L}_{m}=\frac{1}{2}\nabla_\mu\psi\, \nabla^\mu\psi+\nabla_\mu\Phi\,\nabla^\mu \Phi^*+U(\psi, \Phi),\label{lm}\\
&U(\psi, \Phi)=m^2\psi^2\left|\Phi\right|^2+\mu^2\left(\psi^2-v^2\right)^2, \label{pot}
\end{align}
being $\psi$ a self-interacting real scalar field and $\Phi$ a complex scalar field. The real and complex scalar fields are coupled through the coupling constants $m$ and $\mu$. The constant $v$ is the vacuum expectation value of the real scalar field.
The presence of a coupling term in the potential $U(\psi, \Phi)$ generates the effective masses for the real and complex scalar fields. By expanding $U(\psi, \Phi)$ around the minimum $\psi=v$ and $\Phi=0$, we can compute the masses of the scalar fields to be given by $m_\psi=\sqrt{8}\mu\,v$ and $m_\phi=m\,v$. We observe that as $\mu \rightarrow \infty$, the real scalar field becomes infinitely massive and decouples from the complex scalar field. In this limit, the real scalar field tends to $\psi\rightarrow v$, and the E-FLS theory recovers the Einstein-Klein-Gordon model~\cite{Kaup}. On the other hand, as $\mu \rightarrow 0$ the real scalar field becomes massless. The E-FLS star solutions in this limit are the ones that differ most from the mini-boson stars.


The E-FLS field equations can be obtained by varying the action~\eqref{action} with respect to the metric $g_{\mu\nu}$ and the scalar fields $\psi$ and $\Phi$. The variation with respect to the metric leads to the Einstein's equation:
\begin{align}
\label{EFLS}R_{\mu\nu}-\frac{1}{2}R\,g_{\mu\nu}=\kappa^2 T_{\mu\nu},
\end{align}
where the corresponding stress-energy tensor associated to the scalar fields reads
\begin{align}
T_{\mu\nu}=\nabla_\mu\Phi\nabla_\nu\Phi^*+\nabla_\mu\Phi^*\nabla_\nu\Phi+\nabla_\mu\psi\nabla_\nu\psi-\mathcal{L}_m\,g_{\mu\nu}.
\end{align}
On the other hand, by varying the action \eqref{action} with respect to the real and complex scalar fields, we obtain, respectively:
\begin{align}
\label{eom_cfield}&\square\Phi=m^2\,\psi^2\Phi,\\
\label{eom_rfield}&\square\psi=2\psi\left(m^2\left|\Phi\right|^2+2\mu^2\psi^2-2\,v^2\,\mu^2\right),
\end{align}
where $\square\equiv \nabla^\mu\,\nabla_\mu$ is the d'Alembertian operator. We notice that the parameters $v$ and $m$ on Eqs.~\eqref{EFLS}-\eqref{eom_rfield} can be eliminated by performing the following rescaling:
\begin{align}
\nonumber &r=\frac{\tilde{r}}{mv}, \quad \psi=v\tilde{\psi},\quad \phi=v\tilde{\phi},\quad \omega=mv\tilde{\omega},\\ \label{rescal} &\kappa=\frac{\tilde{\kappa}}{v},\quad \mu=m\tilde{\mu},
\end{align}
and the rescaled field equations are equivalent to set $v=m=1$ in Eqs.~\eqref{EFLS}-\eqref{eom_rfield}. In the remainder of this paper, we work with the rescaled field equation, omitting the tildes for the sake of simplicity.

The E-FLS action \eqref{action} possesses a global invariance under a $U(1)$ transformation of the complex scalar field ($\Phi \rightarrow e^{i\alpha}\Phi$),
where $\alpha$ is a constant phase. Hence, by the Noether theorem, a conserved current associated with such symmetry exists and can be expressed as:
\begin{align}
j_{\mu}= i\left(\Phi\,\nabla_{\mu}\Phi^{*}-\Phi^{*}\,\nabla_{\mu}\Phi \right), \label{J}
\end{align}  
such that it has null divergence $\nabla_{\mu}j^{\mu}=0$. The integration of the timelike component of the conserved current over a spacelike hypersurface $\Sigma$ yields a conserved quantity known as the \textit{Noether charge}, namely
\begin{align}
N\equiv \frac{1}{4\pi}\int_{\Sigma} j^{0}\sqrt{-g}d^3x, \label{Npart}
\end{align} 
which is related to the number of scalar particles that constitute the system.

Let us now explore the solitonic solutions in the context of the E-FLS theory. We consider spherically symmetric solutions to the field equations~\eqref{EFLS}-\eqref{eom_rfield}, and adopt the following ansatz for the metric tensor in Schwarzschild-like coordinates:
\begin{align}
\label{metric}ds^2=-e^{\Gamma(r)}dt^2+e^{\Lambda(r)}dr^2+r^2\left(d\theta^2+\sin^2\theta\,d\varphi^2\right),
\end{align}
where $\Gamma(r)$ and $\Lambda(r)$ are two real metric functions that depend solely on the radial coordinate. 
The corresponding ansatzes for the spherically symmetric real and complex scalar fields are
\begin{align}
\label{r_field}&\psi=\psi(r),\\
\label{c_field}&\Phi=\phi(r)e^{i\omega\,t},
\end{align}
respectively, where $\omega>0$ is a real parameter associated with the oscillation frequency of the complex scalar field. Inserting the ansatzes \eqref{metric}-\eqref{c_field} into the E-FLS field equations \eqref{EFLS}-\eqref{eom_rfield}, we obtain the following set of coupled ordinary differential equations (ODEs):
\begin{align}
\nonumber &\frac{e^{-\Lambda}\,\Lambda'}{r}+\frac{\left(1-e^{-\Lambda}\right)}{r^2}=\kappa^2\left[\omega^2\,e^{-\Gamma}\phi^2 \right. \\
\label{ode_gtt} & \left. +e^{-\Lambda}\phi'^2+\frac{e^{-\Lambda}\,\psi'^2}{2}+\psi^2\phi^2+\mu^2\left(\psi^2-1\right)^2 \right],\\
\nonumber &\frac{e^{-\Lambda}\,\Gamma'}{r}-\frac{\left(1-e^{-\Lambda}\right)}{r^2}=\kappa^2\left[-\omega^2\,e^{-\Gamma}\phi^2 \right.\\
&\left. -e^{-\Lambda}\phi'^2-\frac{e^{-\Lambda}\,\psi'^2}{2}-\psi^2\phi^2+\mu^2\left(\psi^2-1\right)^2 \right],\\
&\phi''+\left(\frac{2}{r}+\frac{\Gamma'-\Lambda'}{2}\right)\phi'+e^\Lambda\left(\omega^2 e^{-\Gamma}-\psi^2\right)\phi=0,\\
\nonumber &\psi''+\left(\frac{2}{r}+\frac{\Gamma'-\Lambda'}{2}\right)\psi'+4\mu^2e^{\Lambda}\psi^3\\
\label{psi} &-2\,\left(2\,\mu^2+\phi^2 \right)e^{\Lambda}\psi=0,
\end{align}
where the primes denote differentiation with respect to the radial coordinate $r$.

In order to obtain boson star solutions in the E-FLS theory, we must impose appropriate boundary conditions for Eqs.~\eqref{ode_gtt}-\eqref{psi}, as we outline now. We study configurations that are both regular at the origin $(r=0)$ and asymptotically flat far away from the center ($r=\infty$). The condition of regularity of the E-FLS field equations at the origin implies that
\begin{align}
\label{bc1}\Lambda'(r=0)=0, \quad \phi'(r=0)=0, \quad \psi'(r=0)=0.
\end{align}
Moreover, the asymptotically flatness condition leads to the following boundary conditions:
\begin{align}
\label{bc2}\Gamma(r=\infty)=0, \quad \phi(r=\infty)=0, \quad \psi(r=\infty)=1.
\end{align}
For a given value of the scalar field at the origin $\phi(0)\equiv \phi_0$, the system of equations ~\eqref{ode_gtt}-\eqref{psi}, subjected to the boundary conditions \eqref{bc1}-\eqref{bc2}, defines an eigenvalue problem for the frequency $\omega$. For a given $\phi_0$, there exists an infinite countable set of eigenfrequencies $\lbrace \omega_n \rbrace$ that solves the eigenvalue problem described above. The integer $n$ labels the number of nodes of the complex scalar field. The excited states, characterized by $n\geq 1$, are generically unstable~\cite{LeePang}. Therefore, we study only the solutions with $n=0$. In the next section, we describe the numerical method used to compute the nodeless solutions and present a selection of our numerical results.

\section{Results}
\label{numerical_sol}

\subsection{The numerical method}

In order to solve numerically the two-point boundary value problem described in Sec.~\ref{sec:model}, we used the FORTRAN package entitled COLSYS (Collocation for Systems) originally proposed in Ref.~\cite{colsys}. The COLSYS package employs a spline collocation method at Gaussian points for solving boundary value problems associated with ODEs. The numerical solutions can be obtained by providing an initial guess which is close to the actual solution of the problem. Starting with large values of $\mu$, we use the uncharged mini-boson star solution as our initial guess, since the E-FLS model recovers the EKG model for $\mu\rightarrow \infty$. The mini-boson star initial guess was constructed using a standard shooting method~\cite{Cpp}. Once we obtain the solution in the large $\mu$ limit, we gradually decrease the value of this parameter until $\mu=0$. Since the field equations are invariant under the transformation $\mu\rightarrow -\mu$, it is not necessary to compute numerical solutions for negative values of $\mu$. Thus, we were able to obtain a full scan of the solution space within the E-FLS theory. In this paper, we employed our numerical method using a grid of $2000$ points and a relative accuracy of the order $10^{-11}$.

\subsection{E-FLS stars}

\begin{figure*}
\includegraphics[scale=0.4]{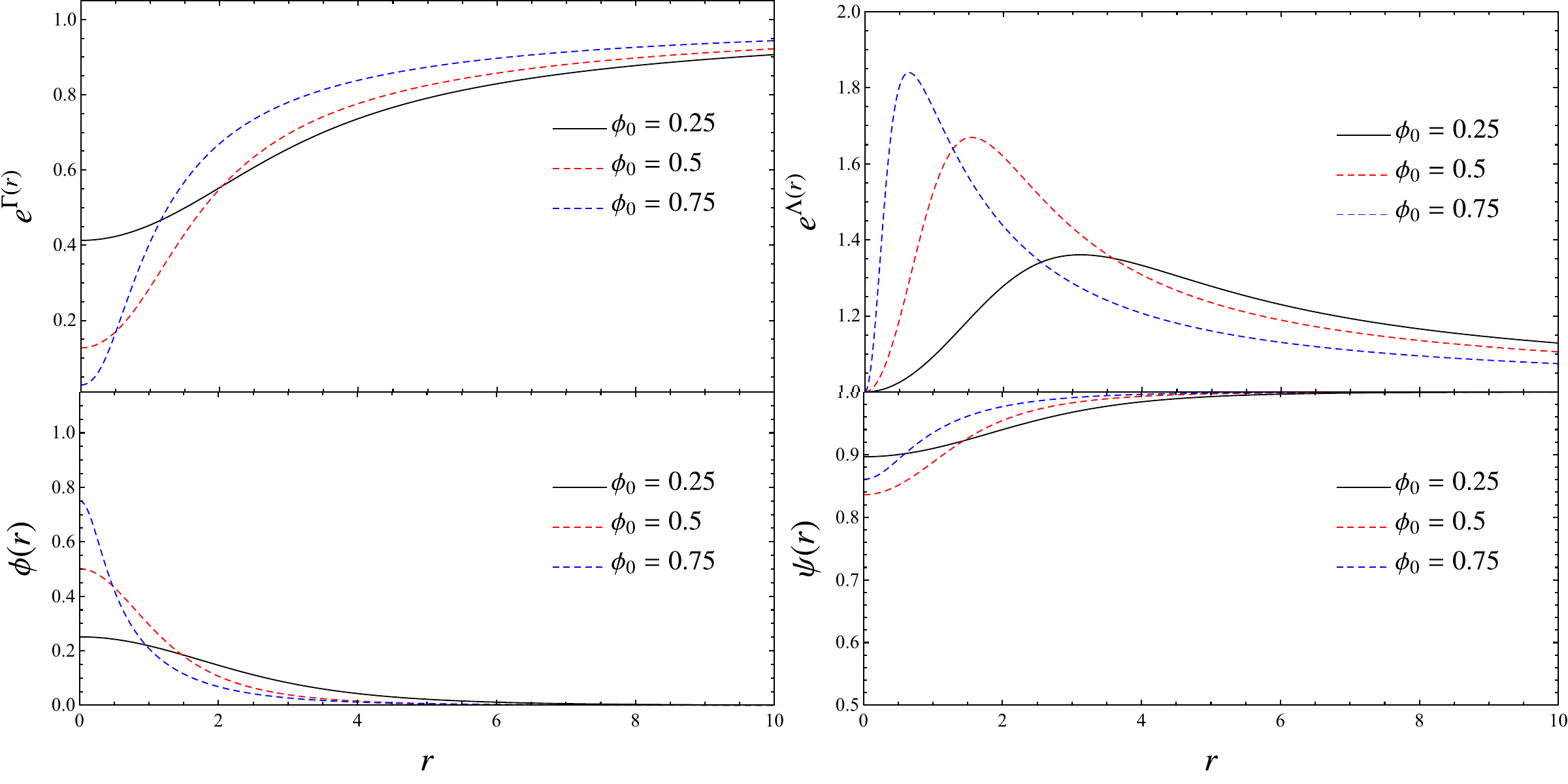}
\caption{
The numerical solutions for E-FLS stars as a function of the radial coordinate. In the top left panel, we display the metric function $e^{\Gamma(r)}$, while in the top right panel we show the metric function $e^{\Lambda(r)}$. In the bottom left and right panels, we show the complex scalar field and real scalar field, respectively. In this figure, we have chosen ${\mu}=0.2$ and displayed three distinct solutions for E-FLS stars with ${\phi}_0 = (0.25, 0.5, 0.75)$.}
\label{fig_1}
\end{figure*}

Let us now present a selection of our numerical results for boson star solution in the E-FLS theory, termed as E-FLS stars in the remainder of this work, for simplicity. We computed the numerical solutions for several values of the parameters $\phi_0$ and $\mu$, in order to fully describe the solution space in this theory. As an example, we show in Fig.~\ref{fig_1}, the numerical solutions for the metric functions $e^{\Gamma(r)}$ and $e^{\Lambda(r)}$, the complex scalar field $\phi(r)$ and the real scalar field $\psi(r)$ in terms of the radial coordinate. In Fig.~\ref{fig_1}, we fixed the value $\mu=0.2$, and varied the value of complex scalar field at the origin. For small values of $\phi_0$, the solutions approach the Newtonian regime, where the star solutions are less compact, and $\omega \approx 1$. As we increase the value of $\phi_0$, the stars become more compact, and relativistic effects tend to dominate the solution, for instance, light rings can emerge in pairs~\cite{Cunha2017} (see Sec.~\ref{Null_geo_sec}). We notice that the E-FLS stars exist in a range of frequency $\omega_{\min}\leq\omega\leq\omega_{\rm max}$, where the upper bound is the Newtonian limit $\omega_{\rm max}=1$ and the lower bound $\omega_{\min}$ depends on the specific value of the FLS parameter $\mu$.


\begin{figure*}
\centering
\subfloat{
\includegraphics[scale=0.69]{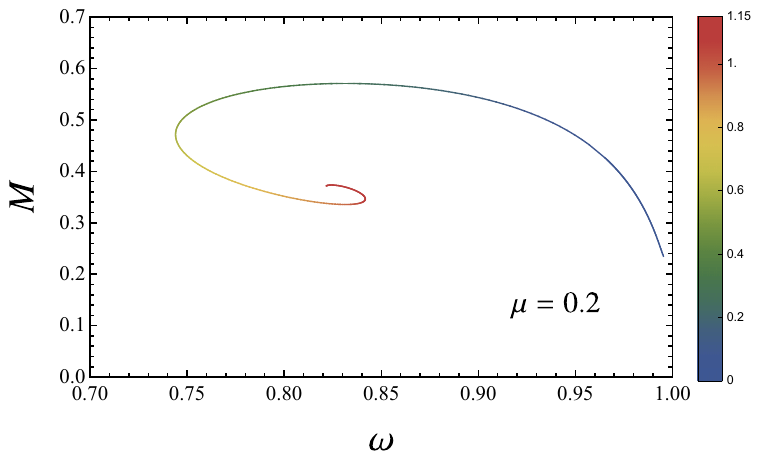}
}
\subfloat{
\includegraphics[scale=0.38]{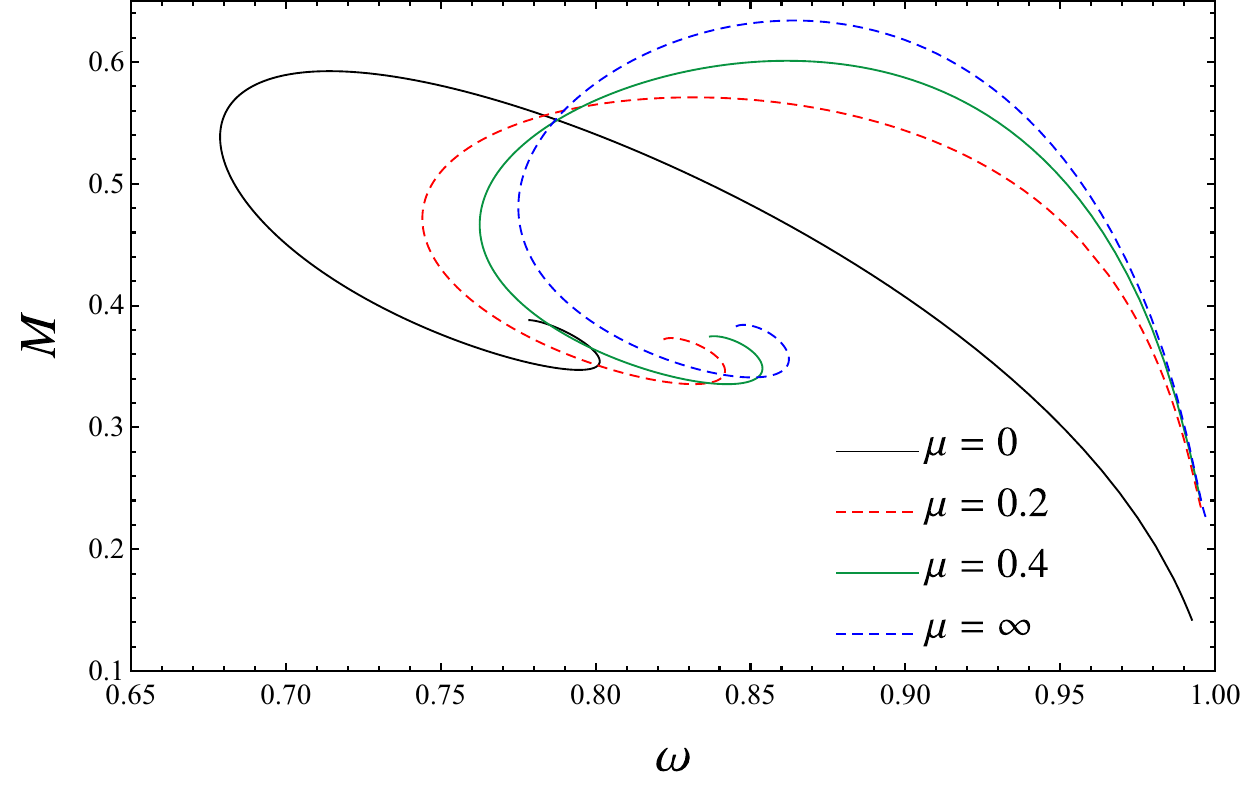}
}
\caption{Left panel: The total mass $M$ for E-FLS stars, with ${\mu}=0.2$, as a function of the normalized frequency ${\omega}$. Each point in this panel represents an E-FLS star solution with a given $\phi_0$. The values of $\phi_0$ are depicted using a color map, with the blue color indicating the lowest value and the red color indicating the highest value. Right panel: The total mass $M$ for E-FLS stars, with different values of $\mu$, as a function of the normalized frequency ${\omega}$. 
}
\label{f2}
\end{figure*}

For a spherically symmetric spacetime, as it is in our case, we can define the mass function $\mathcal{M}(r)$, as given by
\begin{equation}
\mathcal{M}(r) = \frac{r}{2}\left(1- e^{-\Lambda(r)}\right). \label{mass}
\end{equation}
We compute the ADM mass of the E-FLS stars by taking the limit $M\equiv\mathcal{M}(r \rightarrow \infty)$. 
In Fig.~\ref{f2}, we show the ADM mass for E-FLS stars as a function of the oscillation frequency. In the left panel of Fig.~\ref{f2}, we fix the value ${\mu}=0.2$ and show the value of $\phi_0$ as a color map, with the blue (red) color representing the lowest (highest) value. We notice that for small values of $\phi_0$, the Newtonian limit is approached, being the oscillation frequency close to 1. As we increase the value of the complex scalar field at the origin, the E-FLS star becomes more massive, reaching a maximum value at ${\omega}\approx 0.83$. In the right panel of Fig.~\ref{f2}, we show the ADM mass as a function of the oscillation frequency for different values of $\mu$. It can be noticed that the maximum mass depends on the FLS parameter $\mu$. There are essentially two distinct cases: (i) For $\mu\neq 0$, the maximum mass decreases as we decrease the value of $\mu$, being the mini-boson star ($\mu\rightarrow \infty$) case the highest maximum mass. (ii) For $\mu=0$, we notice that the maximum mass increases in comparison to low values of $\mu\neq 0$. Moreover, for the case $\mu=0$, the minimum frequency $\omega_{\min}$ is smaller, in comparison to the case $\mu \neq 0$.
In order to explain the distinct behavior for $\mu=0$, we recall that the mass of the real scalar field, arising due to the interaction with the complex scalar field, is given by $m_\psi=\sqrt{8}\mu\,v$. Hence, for $\mu=0$, the real scalar field becomes massless. In Refs.~\cite{Rubakov2011,Loiko2018}, it was proved that for $\mu\neq 0$ the real scalar field approaches the asymptotic value as $\psi \sim 1-e^{-\sqrt{\mu^2-\omega^2}r}$, while for $\mu=0$ the real massless scalar field decays asymptotically as
\begin{align}
\psi \sim 1-\frac{C}{r}+\mathcal{O}(r^{-2}),
\end{align}
which is akin to the Coulomb asymptotic decay. Therefore, the interaction is long-ranged in the vanishing $\mu$ limit, what explains the distinct behavior in this limit.

In general, the maximum mass solution in a mass-frequency diagram marks the
onset of the instability regime~\cite{LeePang}. The range of solutions, spanning from the Newtonian limit to the maximum mass solution, constitutes the stable branch. A remarkable characteristic of the E-FLS stars is that the family of solutions with $\mu=0$ exhibits a significantly larger stable branch, in comparison to the $\mu\neq 0$ case. Hence, the E-FLS stars are more prone to be stable, as we vary the oscillation frequency, in comparison to the mini-boson stars.

\begin{figure}[h!]
\includegraphics[scale=0.4]{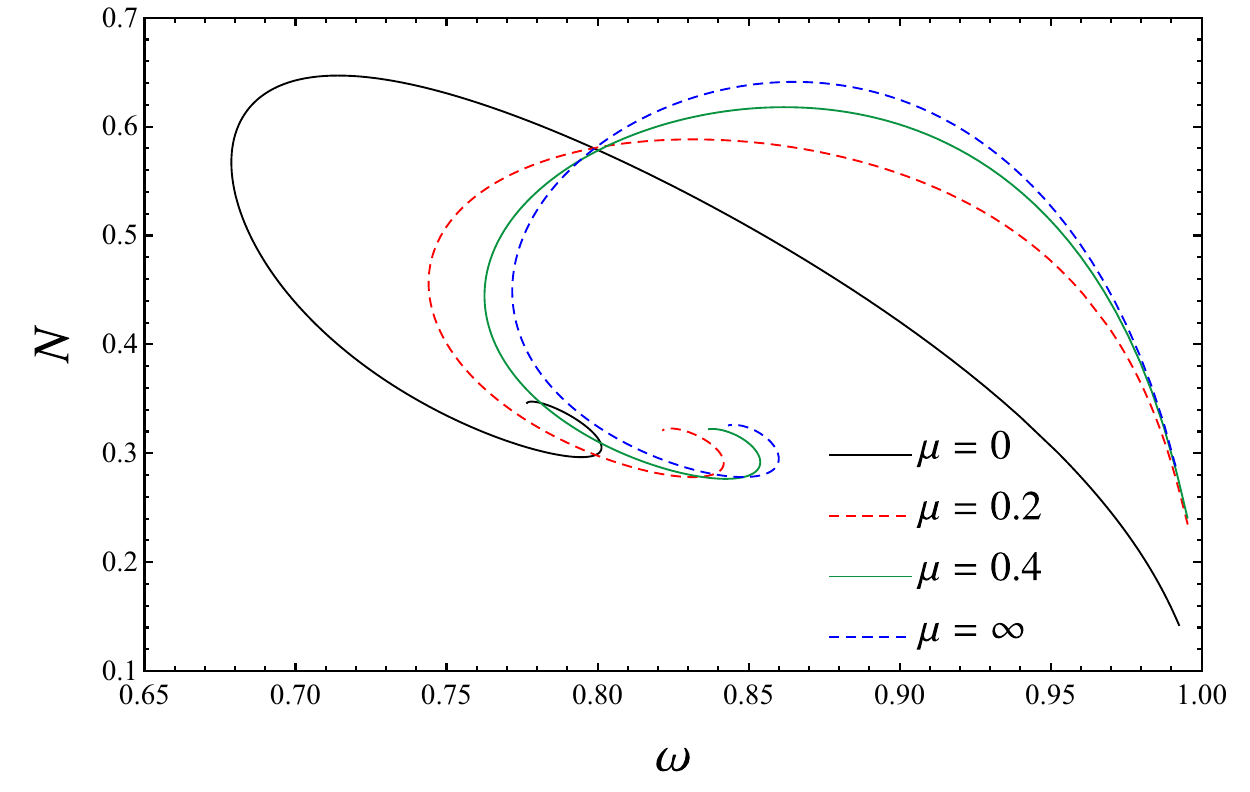}
\caption{The Noether charge for E-FLS stars as a function of the normalized frequency ${\omega}$, for distinct values of the FLS parameter $\mu$. }
\label{Nw}
\end{figure}

\begin{figure*}
\subfloat[$\mu=0$]{
\includegraphics[scale=0.55]{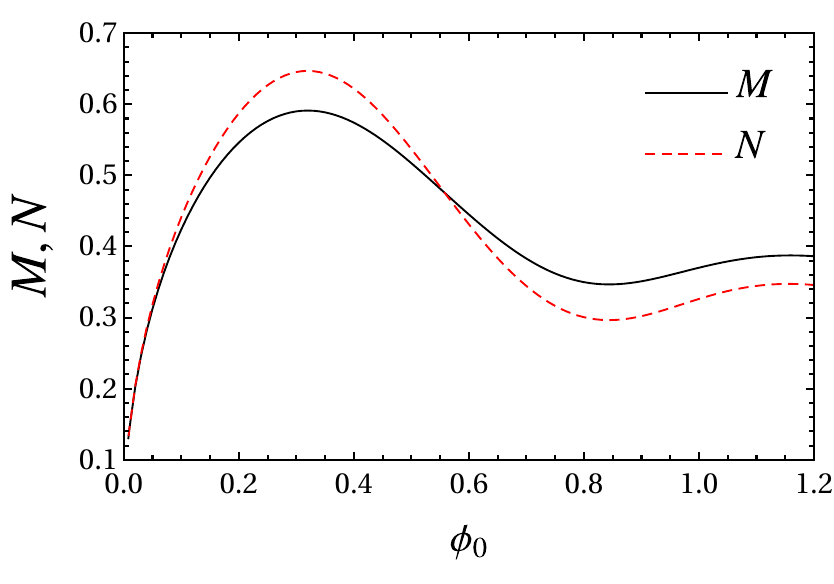}}
\subfloat[$\mu=0.2$]{
\includegraphics[scale=0.55]{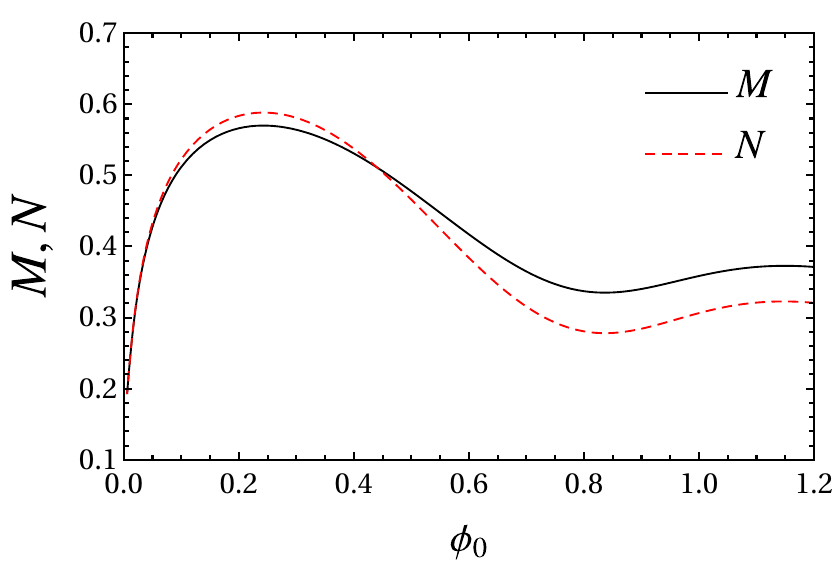}}
\\
\subfloat[$\mu=0.4$]{
\includegraphics[scale=0.55]{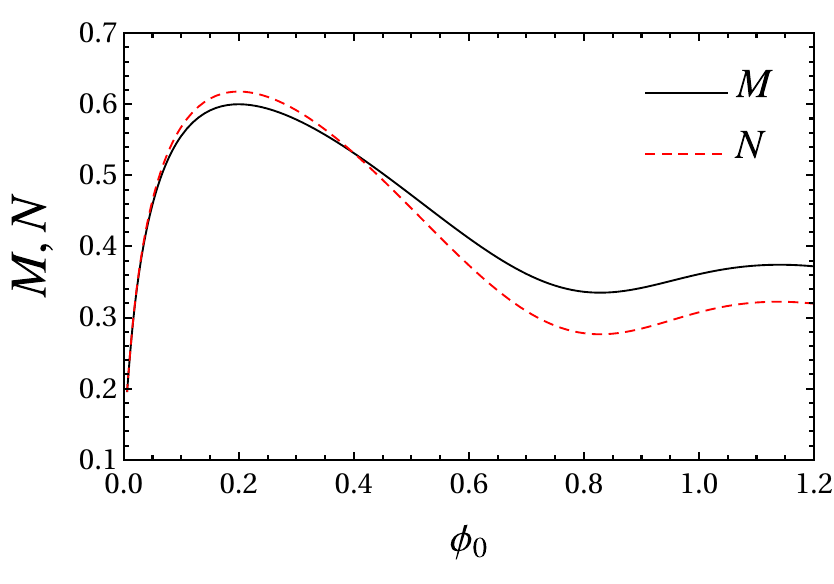}
}
\subfloat[$\mu\rightarrow \infty$]{
\includegraphics[scale=0.55]{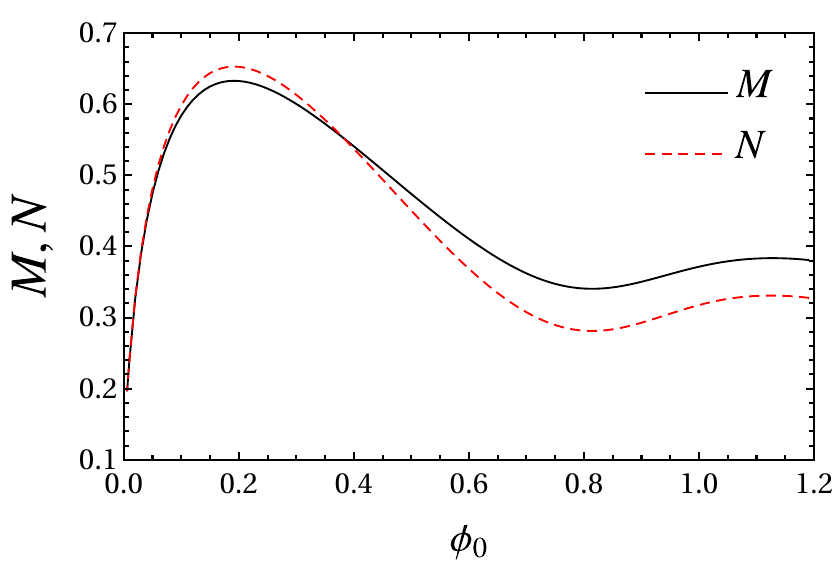}
}

\caption{The total mass $M$ and number of particles $N$ behavior related to the complex scalar field at the origin $\phi_{0}$. We exhibit results for four values of $\mu$, which demonstrate that not only the maximum total mass $M$ decreases for lower values of $\mu$, but also the total number of particles $N$. The set of solutions in which $M < N$ are stable and the others are unstable.}
\label{f3}
\end{figure*}

We also computed the Noether charge, given in Eq.~\eqref{Npart} for the family of E-FLS star solutions. In Fig.~\ref{Nw}, we present the Noether charge as a function of the oscillation frequency, for different values of the FLS parameter $\mu$. As we decrease the FLS parameter, the maximum value for the Noether charge also decreases if $\mu\neq 0$. For $\mu=0$, the maximum value for the Noether charge increases in comparison to the case $\mu\neq 0$, similarly to the behavior observed for the ADM mass.  
The relation between the ADM mass and the Noether charge can provide insights into the stability properties of the E-FLS stars. The binding energy of the star is defined as
\begin{align}
E_{b}=M-m\,N,
\end{align}
therefore boson star solutions with negative binding energy ($M<m\,N$) are gravitationally bound, while solutions with positive binding energy ($M>m\,N$) are gravitationally unbound. In Fig. \ref{f3}, we show the comparison between the total ADM mass and number of particles  as a function of ${\omega}$ and for different values of the FLS parameter. The region with gravitationally bound configuration is centered around the maximum values of the Noether charge and the ADM mass. For high values of ${\phi}_0$, the Noether charge becomes smaller than the total ADM mass, hence the solutions become gravitationally unbound. Decreasing the FLS parameter $\mu$ leads to an increasing of the region associated with gravitationally bound configurations. In particular, the case of $\mu=0$ exhibits the largest region with gravitationally bound configurations, as can be seen in panel (a) of Fig.~\ref{f3}. 

\begin{figure}
\subfloat{
\includegraphics[scale=0.4]{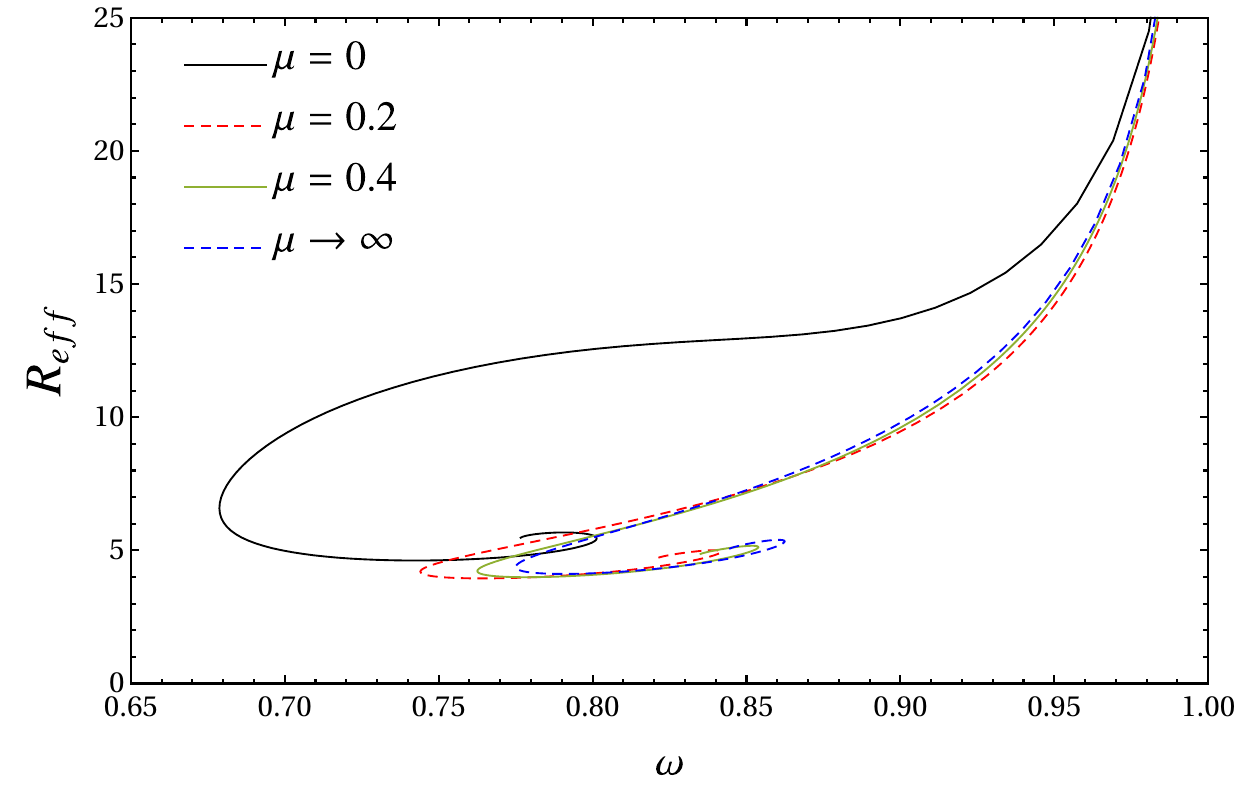}
}\\
\subfloat{
\includegraphics[scale=0.4]{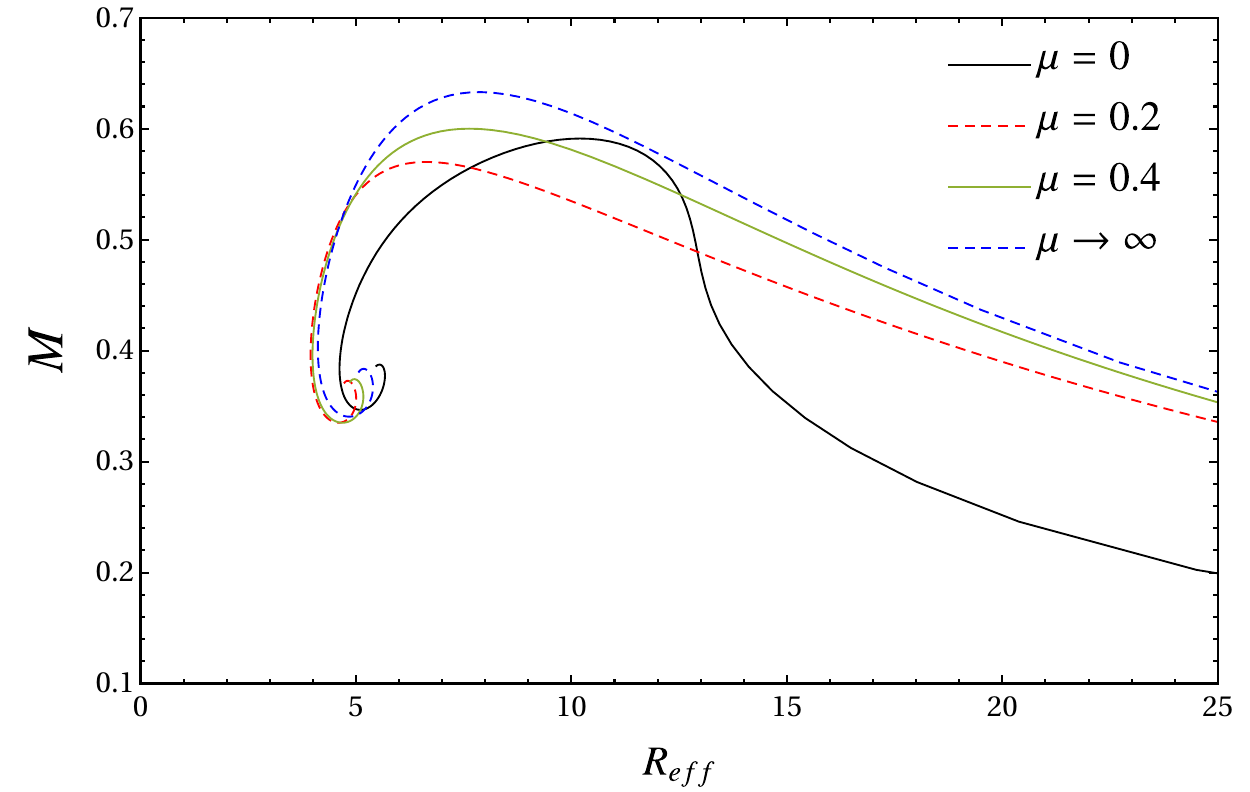}
}
\caption{Top panel: The effective radius $R_{eff}$ as a function of the oscillation frequency for distinct values of $\mu$. Bottom panel: The total mass $M$ of the E-FLS stars in terms of the effective radius $R_{eff}$, for distinct values of the FLS parameter $\mu$. The impact of decreasing the FLS parameter $\mu$ is evident for the $\mu=0$ case.}
\label{f4}
\end{figure}

Besides the ADM mass and the Noether charge, we may also compute an effective radius for the E-FLS stars. In contrast to perfect-fluid stars in GR, boson stars do not posses a well-defined radius since the scalar field is spread up to the spatial infinity. However, for large values of $r$, the scalar field decays exponentially, allowing the definition of an effective radius $R_{eff}$. We note that the definition for $R_{eff}$ is not unique. In the present work, we define the effective radius to be the surface containing $99\%$ of the ADM mass, i.e., $\mathcal{M}(R_{eff})=0.99M$. In the left panel of Fig.~\ref{f4}, we show the effective radius $R_{eff}$ for E-FLS stars as a function of the oscillation frequency ${\omega}$. We observe a slight change in the effective radius for $\mu > 0$. In contrast, a noticeable change in the effective radius is evident for E-FLS stars with $\mu=0$. In particular, the effective radius for $\mu=0$ is considerably larger compared to the other cases, which is in accordance to the fact that for $\mu=0$ the interaction becomes long ranged. In the bottom panel of Fig.~\ref{f4}, we plot the total ADM mass as a function of the effective radius for different values of $\mu$. The noticeable result is observed once more for $\mu=0$, where we note, for instance, that the maximum mass occurs at a larger effective radius compared to the other cases.

\begin{figure}
\includegraphics[scale=0.4]{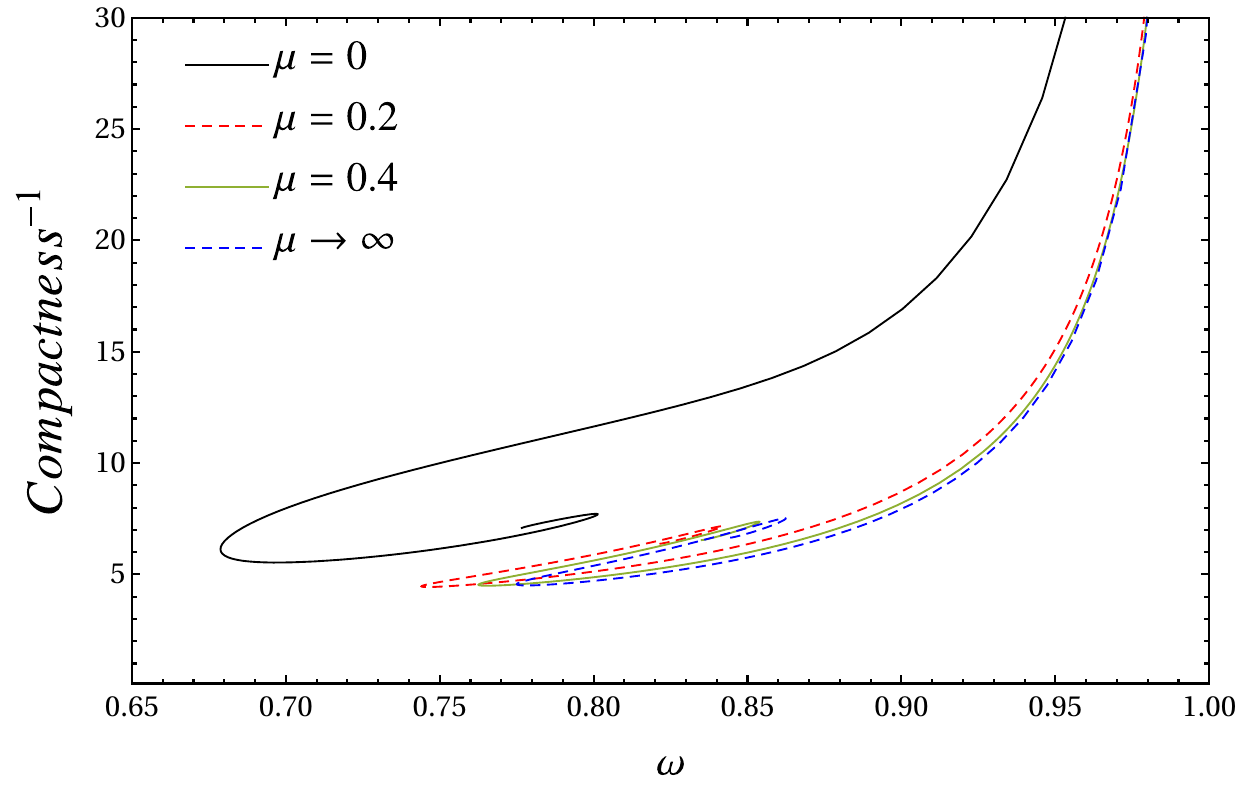}
\caption{The inverse compactness as a function of the oscillation frequency, for distinct values of the FLS parameter $\mu$. We notice that the impact of decreasing the FLS parameter $\mu$ is evident for the inverse compactness in the $\mu=0$ case. }
\label{compac}
\end{figure}

These results for the ADM mass and effective radius suggest that E-FLS stars with $\mu=0$ are less compact than the cases with $\mu>0$. In order to investigate the compactness of the E-FLS stars, we can define the concept of inverse compactness, as given by:
\begin{equation}
\text{Compactness}^{-1} = \frac{R_{eff}}{2M_{eff}}, \label{compact}
\end{equation}
where $M_{eff}$ is the effective mass of the star that represents $99\%$ of its total ADM mass, i.e. $M_{eff}\equiv \mathcal{M}(R_{eff})$. For extremely compact objects, such as black holes, we have $\text{Compactness}^{-1}\sim 1$, while less compact objects have higher inverse compactness. 
In Fig.~\ref{compac}, we show the inverse compactness as a function of the oscillation frequency, for different values of $\mu$. For $\mu\neq 0$, we notice a slight variation of the inverse compactness. However, for $\mu=0$ we notice that inverse compactness increases considerably. This result confirms the expectation that E-FLS stars with $\mu=0$ are less compact.

\section{Geodesic Analysis}
\label{Geo_ana}

\subsection{Timelike geodesics in E-FLS stars}
\label{Subsec_timelike}
\begin{figure}[h!]
\includegraphics[scale=0.6]{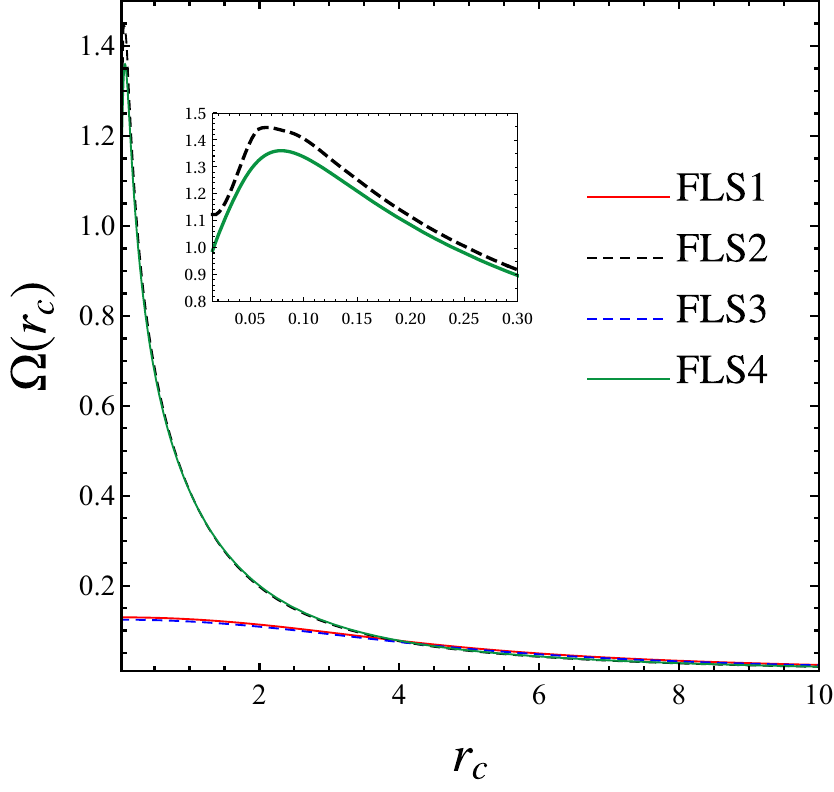}
\caption{The orbital frequency for timelike circular geodesics as a function of the radius $r_c$, for the different E-FLS configurations described in Table~\ref{Sols}. We notice that more compact configurations, for instance FLS2 and FLS4, admit a local maximum for the orbital frequency, as shown in the inset.}
\label{Omega}
\end{figure}

In order to investigate the phenomenological aspects of E-FLS stars, we study the motion of timelike and null geodesics around these compact objects, which is important for the description of accretion disks. Let us first consider timelike geodesics, describing the motion of massive particles around the E-FLS stars, given by the following Lagrangian:
\begin{align}
\label{lagran_timelike}\mathcal{L}_t=g_{\mu\nu}\dot{x}^\mu\dot{x}^\nu=-e^\Gamma\,\dot{t}^2+e^\Lambda\,\dot{r}^2+r^2\dot{\varphi}^2=-1,
\end{align}
where we considered $\theta=\pi/2$, without loss of generality, since the spacetime is spherically symmetric. The overdots denote derivatives with respect to the proper time along the timelike geodesics. We note that the Lagrangian $\mathcal{L}_t$ is independent of $t$ and $\varphi$, hence there are two conserved quantities associated to such symmetries, namely
\begin{align}
\label{energy}&E=e^\Gamma\,\dot{t}, \\
\label{momentum}&L=r^2\dot{\varphi},
\end{align}
such that $E$ and $L$, respectively, represent the energy and angular momentum of the particle. Substituting Eqs.~\eqref{energy}-\eqref{momentum} into the Lagrangian~\eqref{lagran_timelike}, we obtain an equation describing the radial motion, namely:
\begin{align}
\label{rad_eq_timelike}e^{\left( \Gamma+\Lambda\right)}\,\dot{r}^2=E^2-e^\Gamma\left(1+\frac{L^2}{r^2}\right).
\end{align}
The set of circular timelike orbits around E-FLS stars is obtained from the radial equation by imposing $\left. \dot{r} \right|_{r=r_c}=0$ and $\left. \ddot{r} \right|_{r=r_c}=0$. The energy and angular momentum of such orbits at a given radius $r_c$ are given, respectively, by: 
\begin{align}
\label{E-L-timelike}&E_c^2=\left. \frac{2\,e^\Gamma}{2-r\,\Gamma'}\right|_{r=r_c}, \qquad L_c^2=\left. \frac{r^3\,\Gamma'}{2-r\,\Gamma'} \right|_{r=r_c}.
\end{align}
Moreover, using Eqs.~\eqref{energy}-\eqref{momentum} and \eqref{E-L-timelike} we can calculate the orbital frequency of circular geodesics at a radius $r_c$, as follows:
\begin{align}
\label{Omega-timelike}\Omega(r_c)=\left. \frac{d\phi}{dt} \right|_{r=r_c}=\left. \left[\frac{e^\Gamma\,\Gamma'}{2\,r}\right]^\frac{1}{2} \right|_{r=r_c}.
\end{align}
In Fig.~\ref{Omega}, we show the orbital frequency for circular geodesics in E-FLS stars as a function of the radius $r_c$, corresponding to the configurations presented in Table~\ref{Sols}. We notice that the orbital frequency for circular orbits located at a large radius remains unaltered for the different configurations. On the other hand, the orbital frequency close to the origin present distinct behavior for different E-FLS solutions. For instance, the configurations FLS2 and FLS4 present a much larger orbital frequency close to the origin, in comparison to the configurations FLS1 and FLS3. Additionally, the configurations FLS2 and FLS4 present a maximum value for the orbital frequency, which does not occur for the configurations FLS1 and FLS3. 

The properties of timelike circular geodesics studied in the present section are important for the description of accretion disks around E-FLS stars, and consequently for the astrophysical images of such compact objects, as we discuss in details in Sec.~\ref{Shadows}.

\subsection{Null-like geodesics in E-FLS stars}
\label{Null_geo_sec}
The null geodesic trajectories represent the path followed by light in a curved spacetime. The study of null geodesics around E-FLS stars is important due to the connection with the recent images of compact objects at the centers of the M87 and Milky Way galaxies, obtained by the EHT collaboration~\cite{M87_1:2019, M87_2:2019, M87_3:2019, M87_4:2019, M87_5:2019, M87_6:2019, sgra_1:2022, sgra_2:2022, sgra_3:2022, sgra_4:2022, sgra_5:2022, sgra_6:2022}. The Lagrangian for null geodesics around E-FLS stars is given by
\begin{align}
\label{Lagran}2\,\mathcal{L}_{ph}=g_{\mu\nu}\dot{x}^\mu\dot{x}^\nu=-e^\Gamma\,\dot{t}^2+e^\Lambda\,\dot{r}^2+r^2\dot{\varphi}^2=0,
\end{align}
where we considered (again) $\theta=\pi/2$ due to spherical symmetry. Similarly to the timelike geodesic analysis, we may compute the conserved quantities along null geodesics and consequently obtain the radial equation of motion for photons to be given by
\begin{align}
\label{rad_eq}\dot{r}^2+V(r, E, L)=0,
\end{align}
being $V(r, E, L)$ the effective potential for null geodesics, namely
\begin{align}
\label{ef_pot}V(r, L, E)=e^{-\Lambda}\left(\frac{L^2}{r^2}-\frac{E^2}{e^{\Gamma}}\right).
\end{align}
Several properties of the null geodesic trajectories in a E-FLS star can be extracted from the radial equation~\eqref{rad_eq}. In particular, closed circular photon orbits at the equatorial plane, commonly dubbed as light rings (LRs), are described by $\left. \dot{r} \right|_{r=r_{ph}}=0$ and $\left. \ddot{r}\right|_{r=r_{ph}} =0$. The LRs can also be described as critical points of the effective potential:
\begin{align}
\label{LR1} V(r_{ph}, L, E)=0, \quad V'(r_{ph} , L, E)=0.
\end{align}
We observe that $V(r, L, E)$ has the disadvantage of depending on the constants of motion $E$ and $L$. For this reason, we choose to work with a new effective potential $\mathcal{H}(r)$, which depends solely on the radial coordinate and is defined by:
\begin{align}
\label{H_pot} &\mathcal{H}(r)\equiv \frac{e^{\frac{\Gamma}{2}}}{r},\\
&V(r, L, E)=\frac{L^2}{e^{\Gamma+\Lambda}}\left(\mathcal{H}(r)+\frac{1}{b}\right)\left(\mathcal{H}(r)-\frac{1}{b}\right),
\end{align}
where $b\equiv L/E$ is the impact parameter for photons. From Eqs.~\eqref{LR1} and \eqref{H_pot}, we can obtain the conditions for the existence of a LR in terms of the new effective potential $\mathcal{H}(r)$, namely:
\begin{align}
\label{LR_Conds}\mathcal{H}(r)=\frac{1}{b}, \quad \mathcal{H}'(r)=0,
\end{align}
where we notice that LRs are critical points of the new effective potential as well.

\begin{figure}
\includegraphics[scale=0.4]{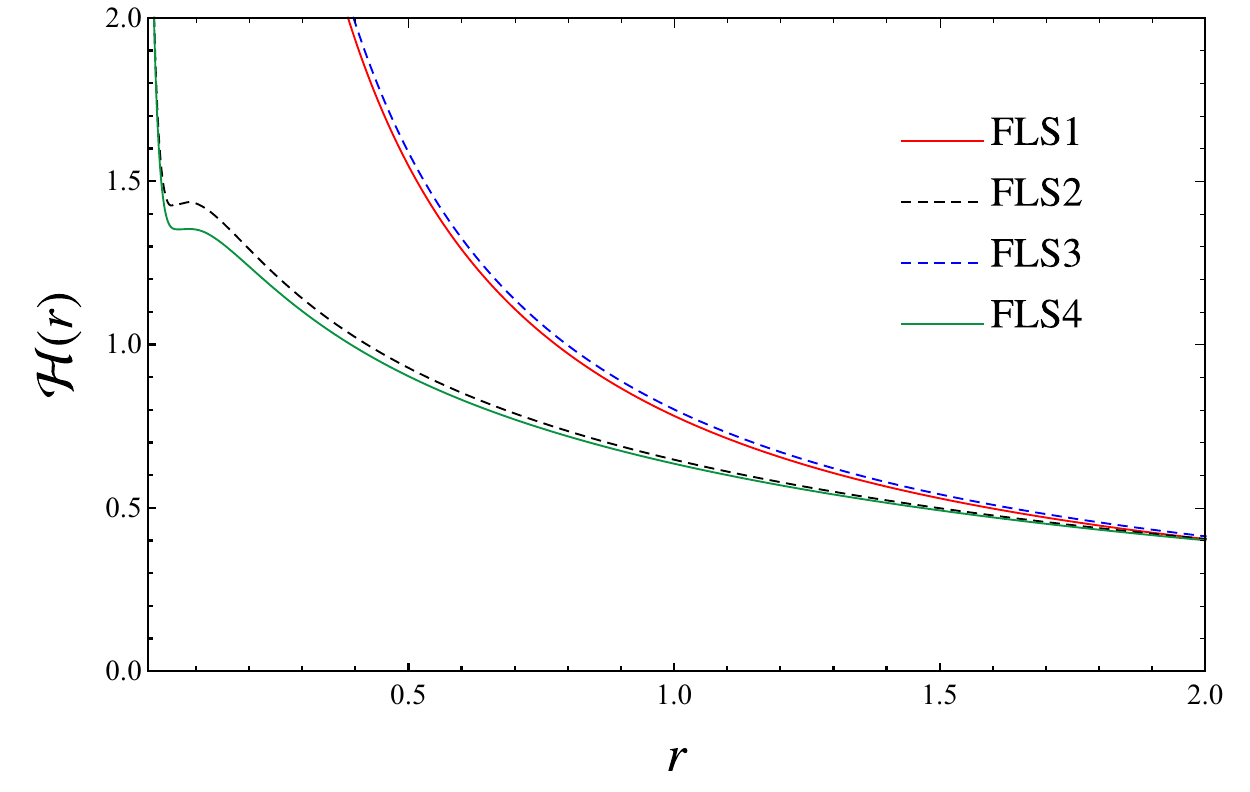}
\caption{The effective potential $\mathcal{H}(r)$ for null geodesics around E-FLS stars, as a function of the radial coordinate, and for the different configurations presented in Table~\ref{Sols}. We notice that, depending on the particular configuration, the star can have two LRs or no LR at all.}
\label{eff_pot}
\end{figure}

In Fig.~\ref{eff_pot}, we show different examples for the effective potential $\mathcal{H}(r)$ as a function of the radial coordinate, computed for the distinct configurations presented in Table~\ref{Sols}. We observe that for the configurations FLS1 and FLS3, there are no light rings at all. This result is expected, as these configurations are close to the Newtonian regime, where E-FLS stars are not sufficiently compact. On the other hand, LRs emerge for the configurations FLS2 and FLS4. We note that they always appear as pairs of unstable/stable LRs, in accordance with the theorems for the conservation of the total topological charge for horizonless compact objects~\cite{Cunha2017, Cunha2020}. Denoting the radial coordinate of the unstable LR as $r_+$ and the radial coordinate of the stable LR as $r_-$, we have that $r_+>r_-$ for any E-FLS star admitting LRs.

\begin{figure*}
  \centering
  \subfloat[FLS1]{\includegraphics[scale=0.35]{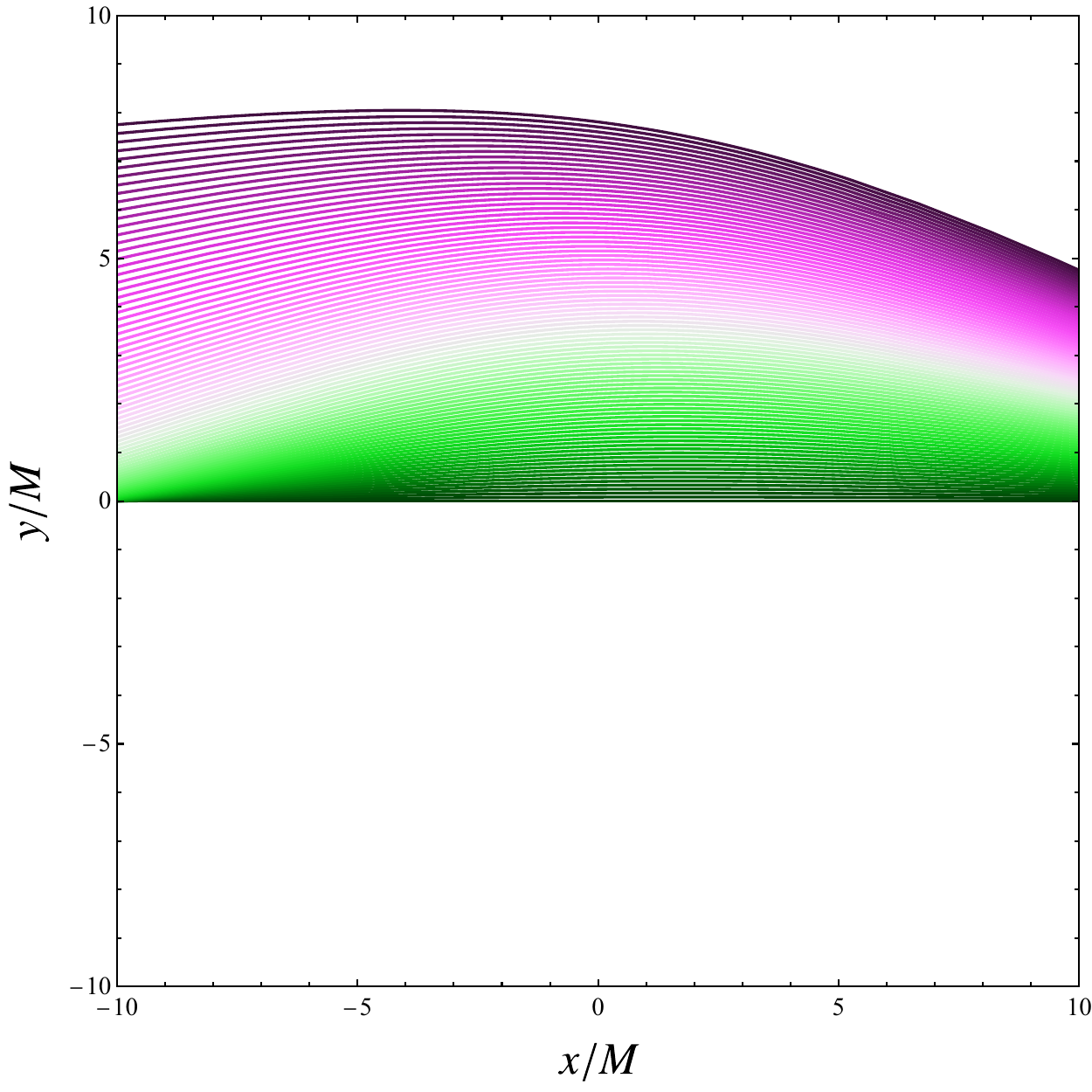}\label{ageo}}
  \subfloat[FLS2]{\includegraphics[scale=0.35]{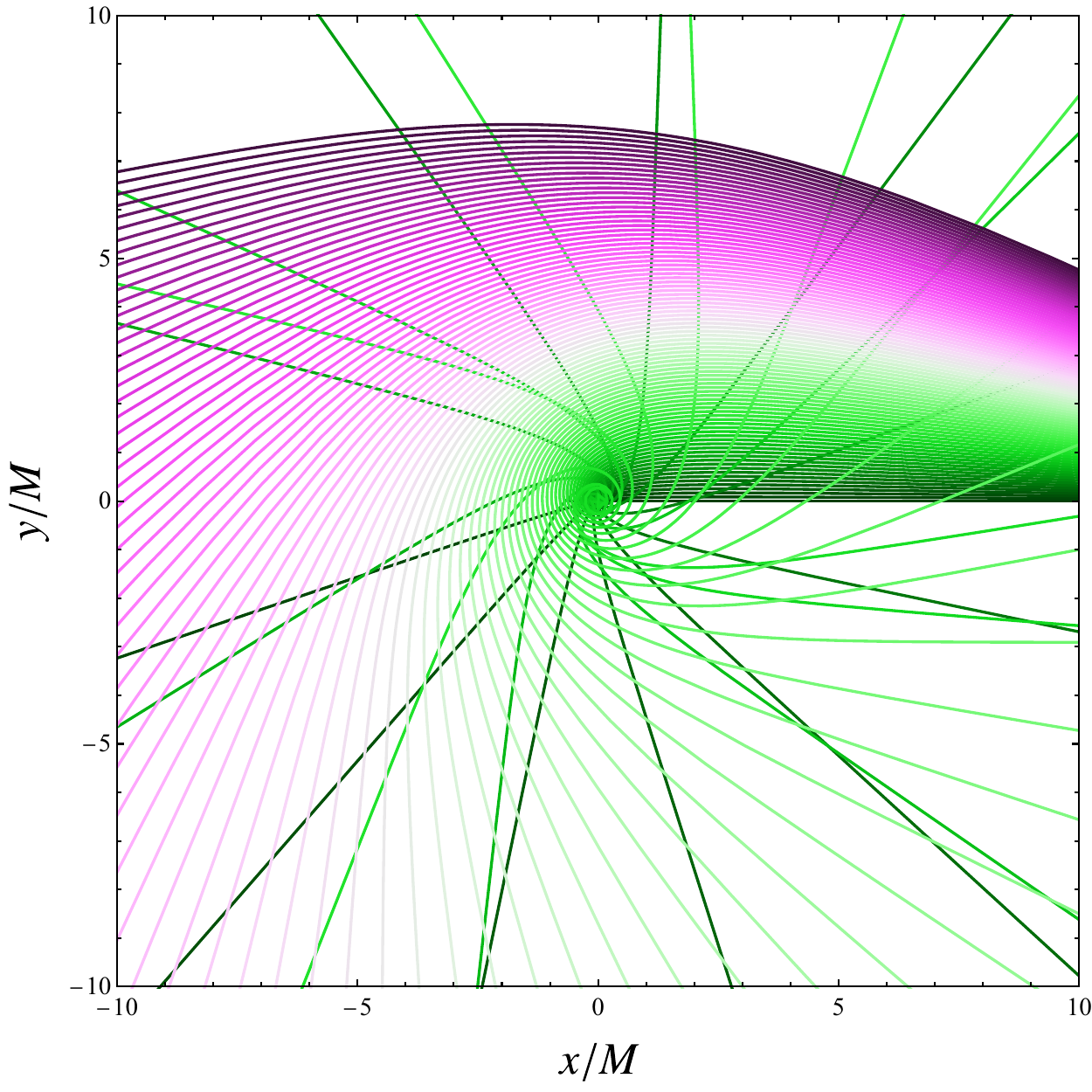}\label{bgeo}}
  \\
  \subfloat[FLS3]{\includegraphics[scale=0.35]{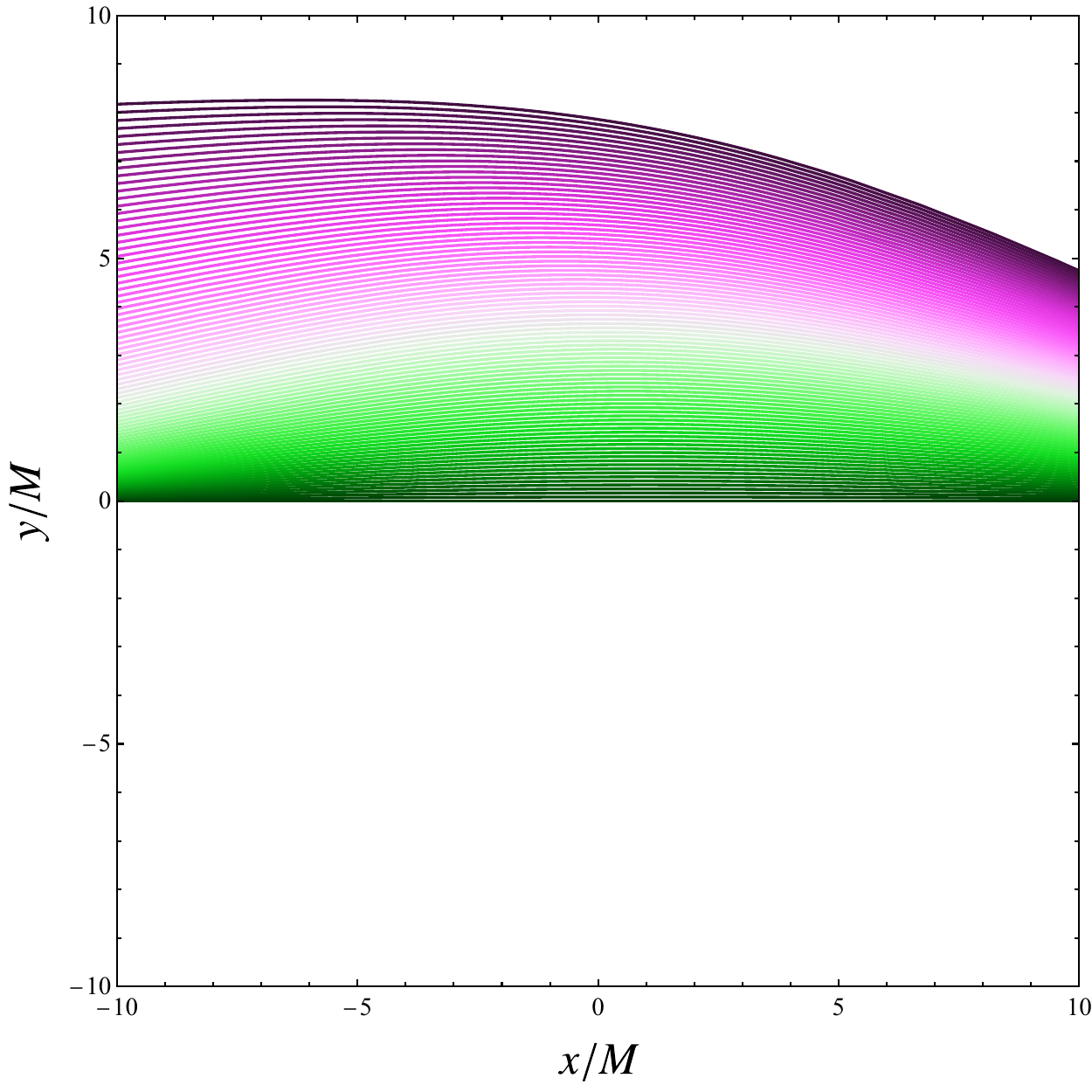}\label{cgeo}}
  \subfloat[FLS4]{\includegraphics[scale=0.35]{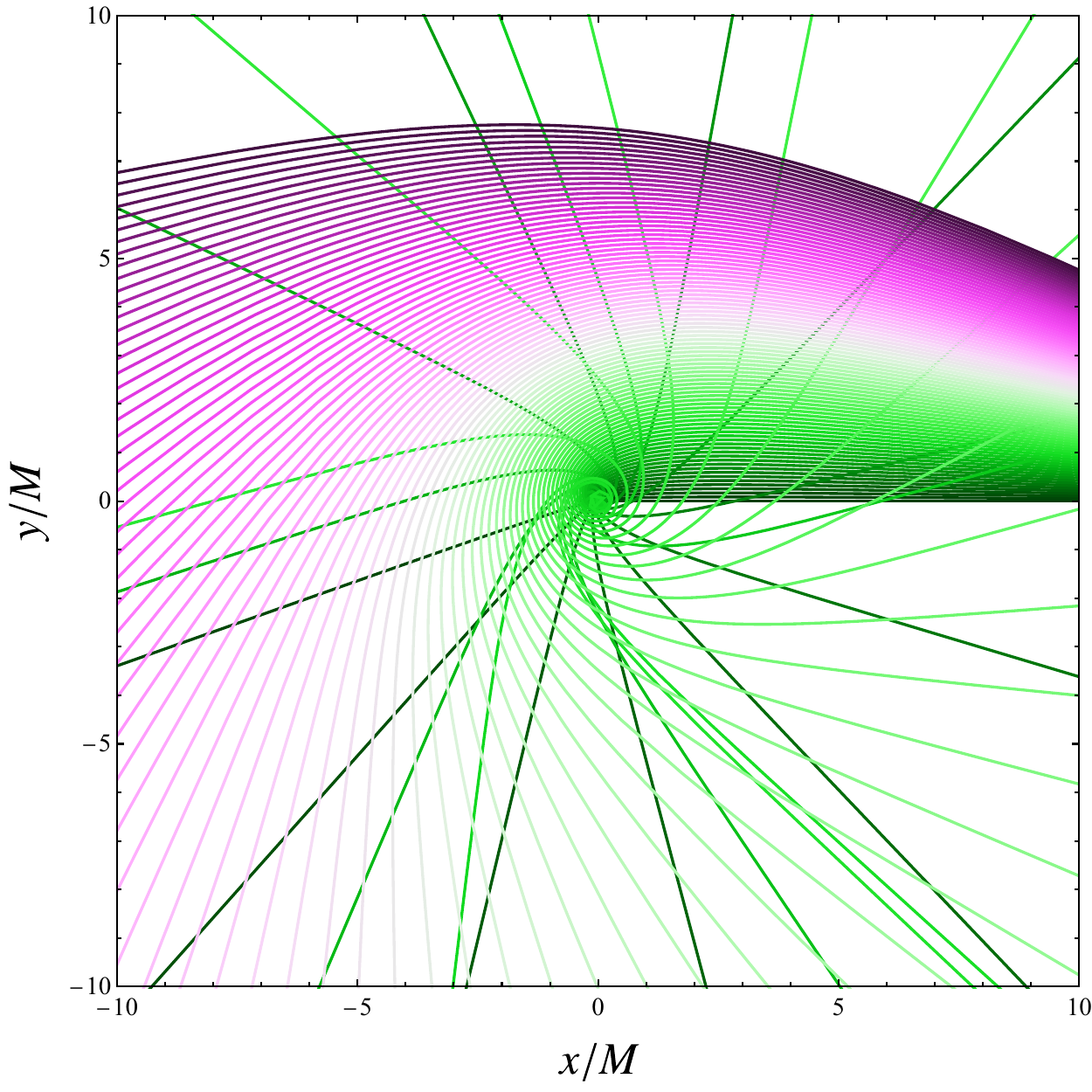}\label{dgeo}}
\caption{The trajectories described by null geodesics with different impact parameters along the equatorial plane of the E-FLS stars. The color scheme represents the different impact parameters for null geodesics. }
\label{geo_traject_FLS}
\end{figure*}

In Fig.~\ref{geo_traject_FLS}, we show the null geodesic trajectories around the different E-FLS stars configurations presented in Table~\ref{Sols}. The null geodesics are restricted to the equatorial plane and have distinct impact parameters. We notice that for the configurations FLS1 and FLS3 the null geodesics are slightly deflected, while for the configurations FLS2 and FLS4 they can be scattered with arbitrarily large angles. These results are in accordance with the fact that the configurations FLS1 and FLS3 are not compact enough, in contrast to the configurations FLS2 and FLS4 which posses light rings.

\begin{table}[htp]
\caption{Four distinct E-FLS star solutions and their properties.}
\centering
\label{Sols}
\begin{ruledtabular}
\begin{tabular}{@{\hspace{0em}}  c @{\hspace{0em}} c @{\hspace{0em}} c  c c c}
Configuration &$\mu$ &  $\omega$& $M$ &Number of LRs & $r_+$\\
\hline
FLS1&\ \  0.2\ \ \   & $0.8873$ &0.5523 & 0  & --\\ \hline
FLS2& \ \ \ 0.2 \ \ \  & $0.8280$ &0.3715 & 2& $0.0794$\\  \hline
FLS3& 0 & 0.8323 & 0.5043 & 0 & --\\  \hline
FLS4& 0 & 0.7828 & 0.3855 &2& $0.0896$
\end{tabular}
\end{ruledtabular}
\end{table}

The existence of such LRs for the configurations FLS2 and FLS4 is a purely relativistic effect and can impact significantly to the visual appearance of the E-FLS stars. In order to investigate the images of these configurations, we discuss in the next section the implementation of the backwards ray-tracing technique for the E-FLS stars surrounded by different accretion disk models.

\section{Astrophysical Images of E-FLS stars}
\label{Shadows}
We found that some E-FLS star configurations admit LRs for given values of the FLS parameter and of the frequency. The presence of LRs can impact significantly the gravitational lensing, and consequently the images of E-FLS stars, as seen by a distant observer. In order to simulate such images, we apply the so-called backwards ray-tracing technique to the E-FLS stars~\cite{Bohn2015, Cunha2018, Cunha20202, HCDL2021, Proca_Star_Shadow, Ivo_Proca_Shadow, Caio_shadows}. The backwards ray-tracing consists in evolving the light rays from the observer's position, backwards in time, until some event is met, for instance, when the light ray intersects the accretion disk or escape to infinity. The implementation of the backwards ray-tracing method involves solving the first order equations of motion for null geodesics
\begin{align}
\label{energy_ph}&\dot{t}=\frac{E}{e^\Gamma},\\
\label{momentum_ph}&\dot{\phi}=\frac{L}{r^2},
\end{align}
coupled to the second order equations of motion for the radial and polar coordinates:
\begin{align}
\label{dotr}\ddot{r}+\Gamma^r_{\ \mu\nu}\dot{x}^\mu\dot{x}^\nu=0,\\
\label{dottheta}\ddot{\theta}+\Gamma^{\theta}_{\ \mu\nu}\dot{x}^\mu\dot{x}^\nu=0,
\end{align}
where $\Gamma^\alpha_{\ \mu\nu}$ denotes the components of the Christoffel symbol associated to the E-FLS star geometry. We note that for the backwards ray-tracing technique, the motion is not assumed to be restricted to the equatorial plane, hence we need to solve also the equation of motion for the polar coordinate~\eqref{dottheta}.

Since we are interested in simulating the images of E-FLS stars surrounded by an emitting accretion disk, we need to solve the so-called radiative transfer equation, coupled to the geodesic equations. Assuming that the beam of photons are unpolarized, the radiative transfer equation is given by~\cite{Lindquist}.

\begin{align}
\label{RTE}\frac{d}{d\lambda}\left(\frac{I_\nu}{\nu^3} \right)=\frac{j_\nu}{\nu^2} - \nu\,\alpha_\nu\,\left(\frac{I_\nu}{\nu^3} \right),
\end{align}
where $I_\nu, j_\nu$ and $\alpha_\nu$ are the specific intensity, emission coefficient and absorption coefficient, respectively. These quantities are measured by an observer comoving with the accretion disk. They are related to the invariant quantities, i.e. observer independent objects, as follows:
\begin{align}
\label{invar_RTE}\mathcal{I}=\frac{I_\nu}{\nu^3}, \quad \eta=\frac{j_\nu}{\nu^2}, \quad \chi=\nu\,\alpha_\nu,
\end{align}
being $\nu$ the frequency of emission. 

In order to obtain the initial conditions for the set of Eqs.~\eqref{energy_ph}-\eqref{momentum_ph} and \eqref{dotr}-\eqref{dottheta}, we apply a procedure akin to the one detailed in Refs.~\cite{Bohn2015, Cunha2018, Cunha20202, HCDL2021}, which we summarize as follows:
\begin{itemize}
\item (i) Since we are propagating the light rays starting at the observer's position, the first set of initial conditions is given by 
$t=0,\ r=r_{obs},\ \theta=\theta_{obs},\ \varphi=0 ;$
\item (ii) A second set of initial conditions can be obtained by computing the initial direction of the light ray, as measured by a static local observer. We introduce a local tetrad associated to the static observer and compute the projection of the 4-momentum of the light ray into the tetrad basis;
\item (iii) Each initial direction, as seen by the local observer, defines a geodesic to be numerically evolved;
\item (iv) Finally, by evolving the geodesic equation coupled to the radiative transfer equation, for a given set of initial directions parametrized by two angles $(\alpha, \beta)$, we produce a map: ($\alpha$, $\beta$) $\mapsto I^{obs}(x,y)$, where $(x, y)$ are two coordinates on the image plane, which is perpendicular to the observer’s line of sight, and $I^{obs}(x,y)$ is the corresponding observed intensity, which represents the astrophysical image of the E-FLS stars. 
\end{itemize}
As the solutions for E-FLS stars are numerical rather than analytical, we have also incorporated an interpolation scheme in our ray-tracing code. The chosen interpolation method is the cubic spline, providing a ${C}^2$ interpolation function. 

We have chosen the four distinct configurations presented in Table~\ref{Sols} for the backward ray-tracing simulation. 
In the next subsections, we explore the astrophysical images for these configurations surrounded by two different models for the accretion disk.

\begin{figure*}
  \centering
  \subfloat[FLS1]{\includegraphics[scale=0.35]{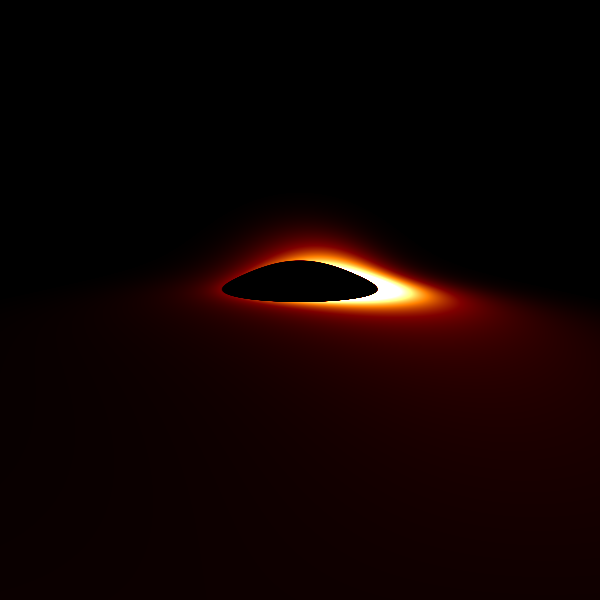}\label{a}}
  \subfloat[FLS2]{\includegraphics[scale=0.35]{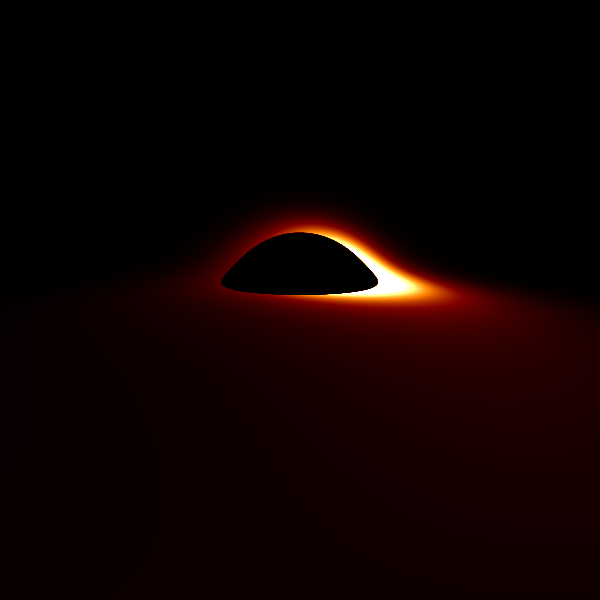}\label{b}}
  \\
  \subfloat[FLS3]{\includegraphics[scale=0.35]{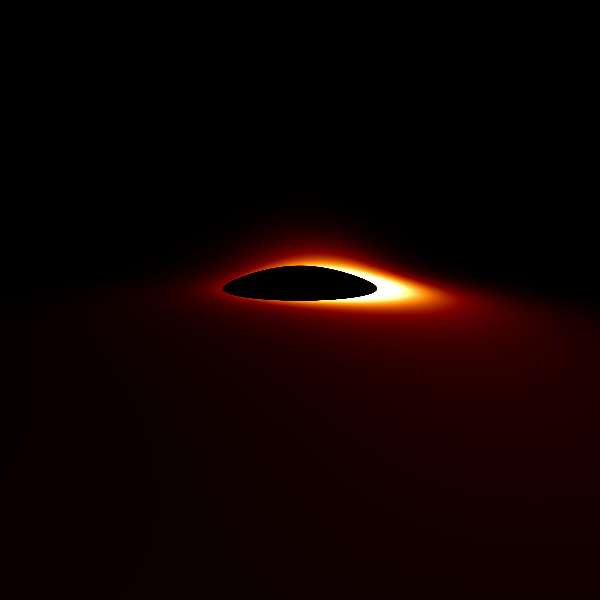}\label{c}}
  \subfloat[FLS4]{\includegraphics[scale=0.35]{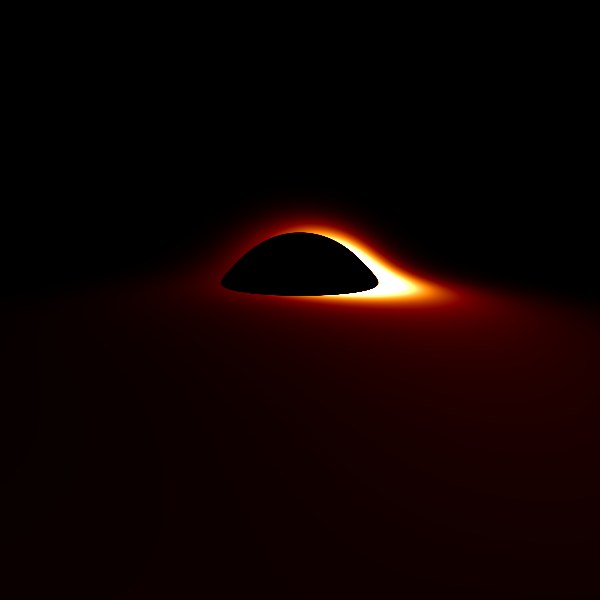}\label{d}}
\caption{The gravitational lensing for the distinct E-FLS star configurations, presented in Table~\ref{Sols}, surrounded by an optically thick accretion disk. In this figure we have positioned the observer at $r_{obs}=20\,M$, $\theta_{\textrm{obs}}=80^\circ$, i.e. the observer is slightly displaced from the equatorial plane.}
\label{Lensing_EFLS}
\end{figure*}

\begin{figure*}
  \centering
  \subfloat[FLS1]{\includegraphics[scale=0.35]{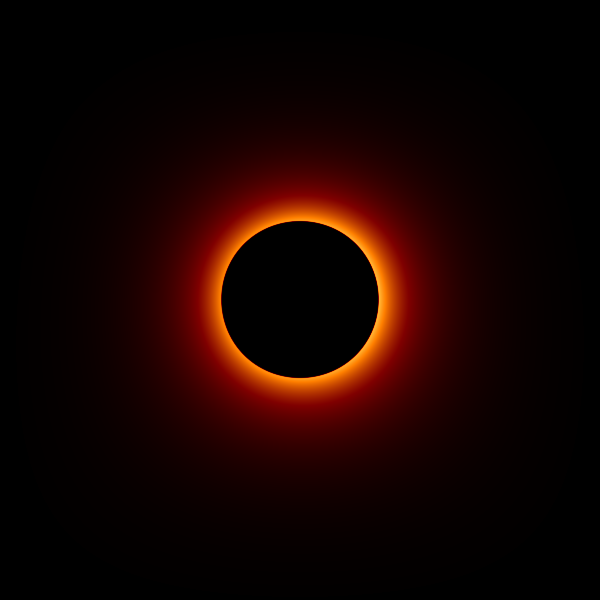}\label{a2}}
  \subfloat[FLS2]{\includegraphics[scale=0.35]{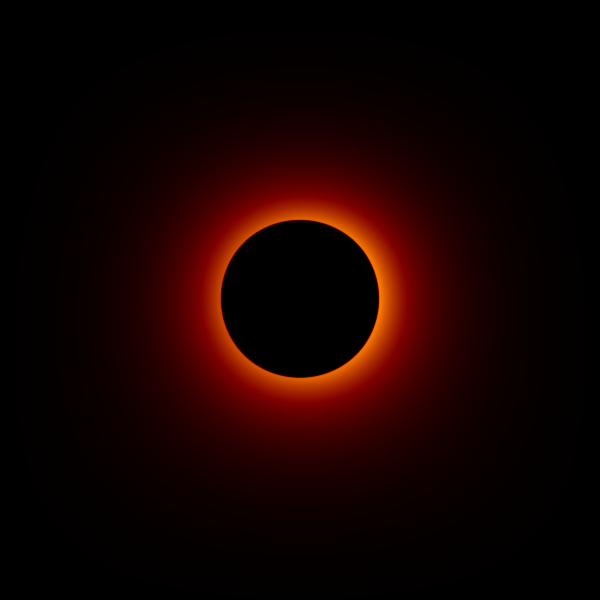}\label{b2}}
  \\
  \subfloat[FLS3]{\includegraphics[scale=0.35]{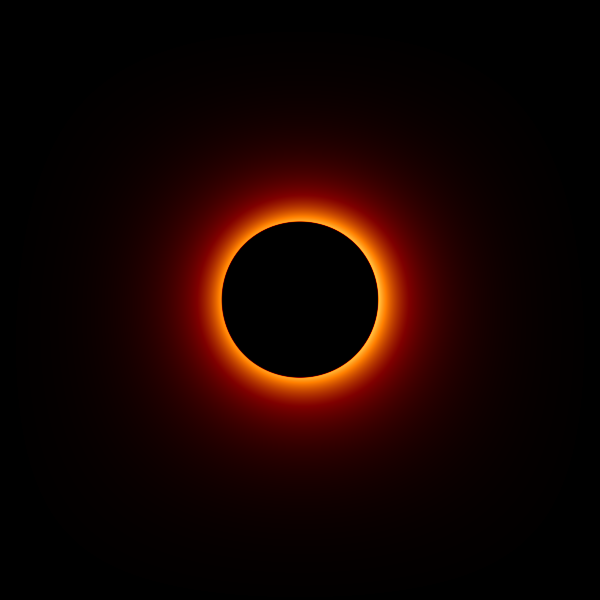}\label{c2}}
  \subfloat[FLS4]{\includegraphics[scale=0.35]{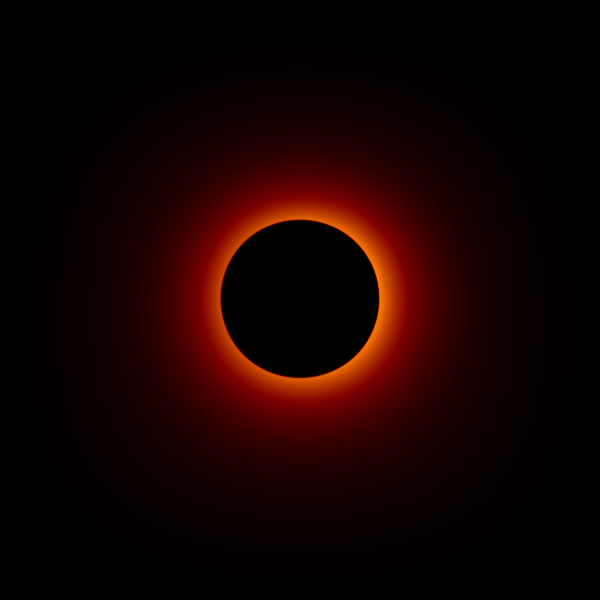}\label{d2}}
\caption{The gravitational lensing for the distinct E-FLS star configurations, presented in Table~\ref{Sols}, surrounded by an optically thick accretion disk. In this figure the observer is placed at $r_{obs}=20\,M$, $\theta_{\textrm{obs}}=5^\circ$, i.e. the observer is close to a face-on observation of the accretion disk.}
\label{Lensing_EFLS_axis}
\end{figure*}

\begin{figure*}
  \centering
  \subfloat[FLS1]{\includegraphics[scale=0.35]{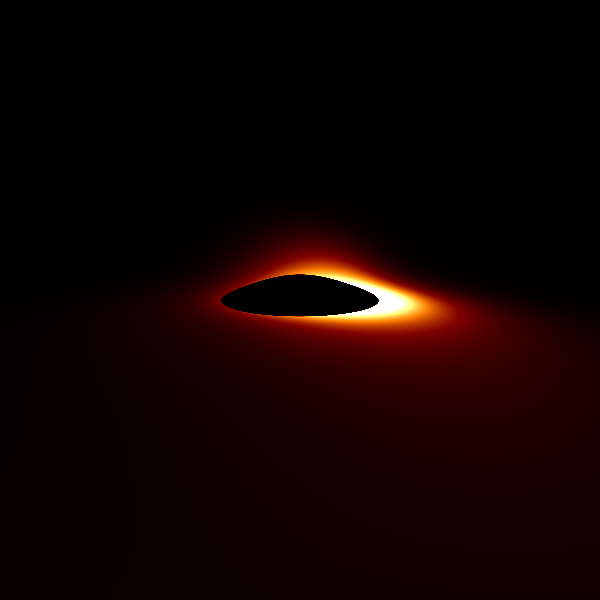}\label{a1}}
  \subfloat[FLS2]{\includegraphics[scale=0.35]{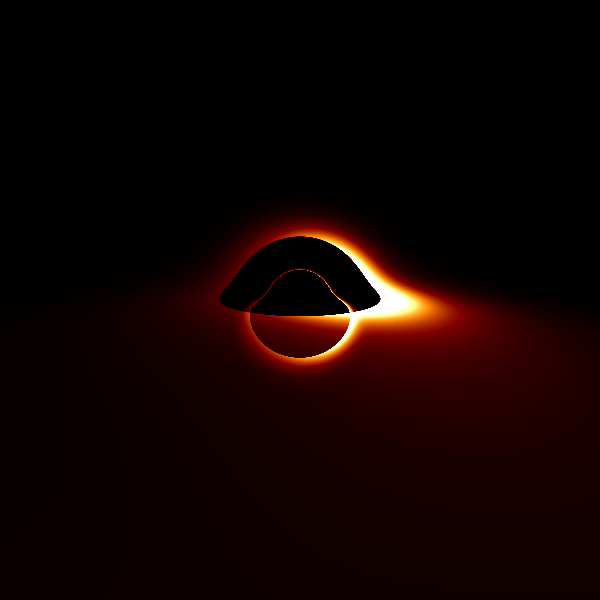}\label{b1}}
  \\
  \subfloat[FLS3]{\includegraphics[scale=0.35]{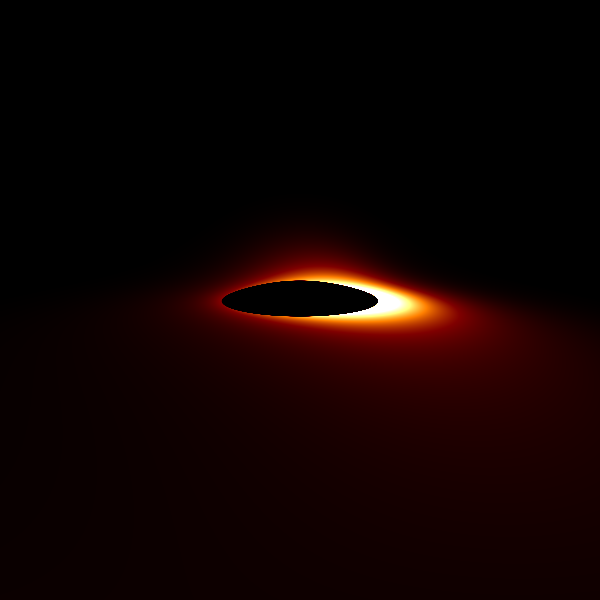}\label{c1}}
  \subfloat[FLS4]{\includegraphics[scale=0.35]{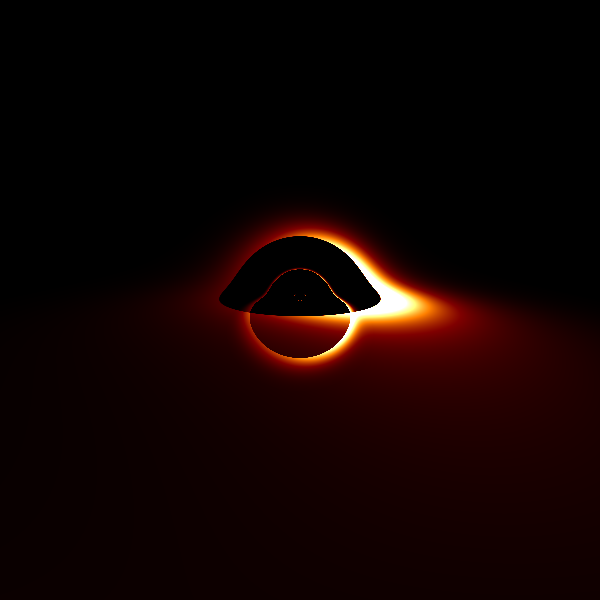}\label{d1}}
\caption{The gravitational lensing for the distinct E-FLS star configurations, presented in Table~\ref{Sols}, surrounded by an optically thin accretion disk. Similarly to Fig~\ref{Lensing_EFLS}, we have positioned the observer at $r_{obs}=20\,M$, $\theta_{\textrm{obs}}=80^\circ$.}
\label{Lensing_EFLS_thin}
\end{figure*}

\begin{figure*}
  \centering
  \subfloat[FLS1]{\includegraphics[scale=0.35]{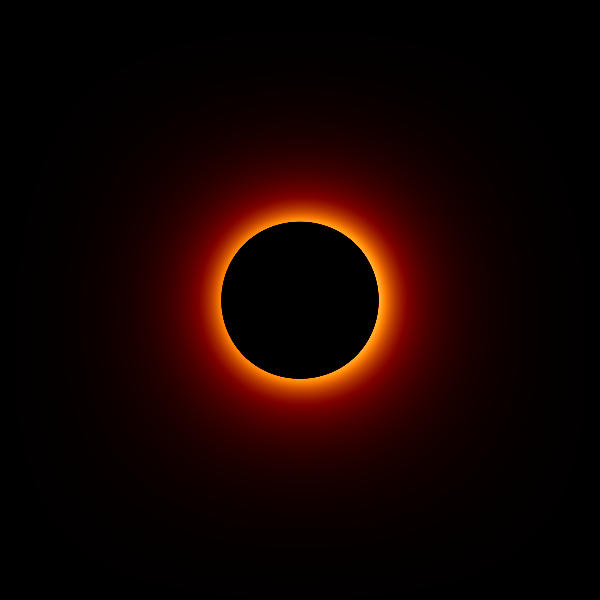}\label{a3}}
  \subfloat[FLS2]{\includegraphics[scale=0.35]{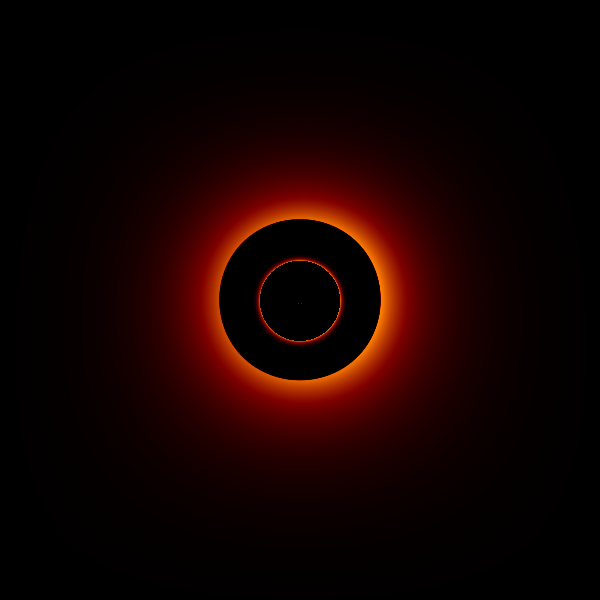}\label{b3}}
  \\
  \subfloat[FLS3]{\includegraphics[scale=0.35]{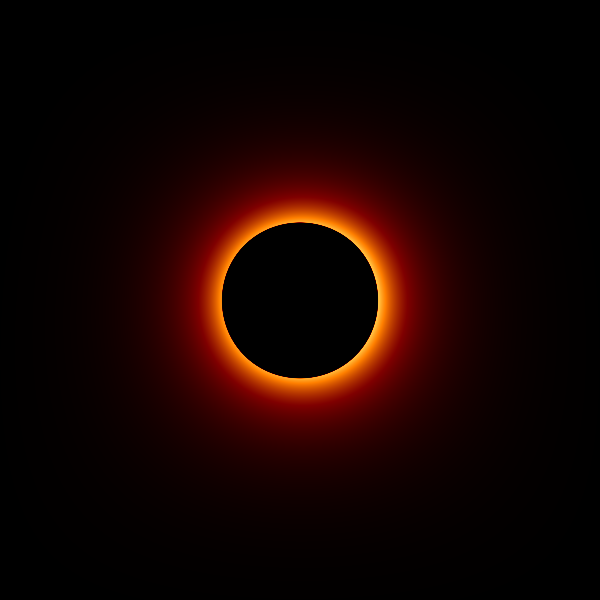}\label{c3}}
  \subfloat[FLS4]{\includegraphics[scale=0.35]{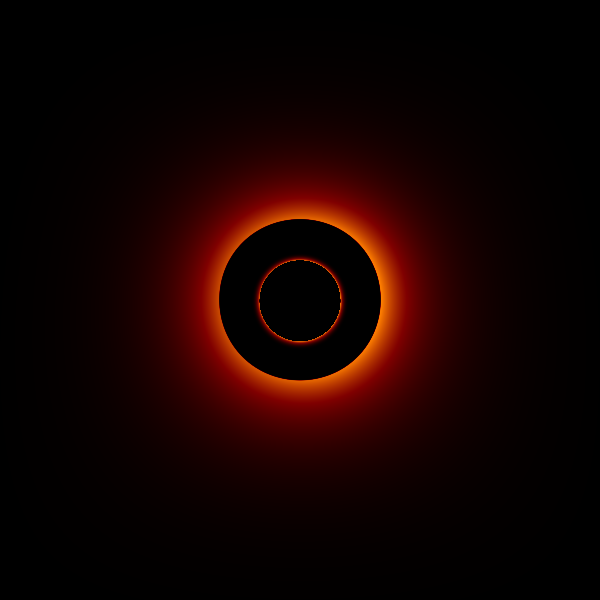}\label{d3}}
\caption{The gravitational lensing for the distinct E-FLS star configurations, presented in Table~\ref{Sols}, surrounded by an optically thin accretion disk. Similarly to Fig~\ref{Lensing_EFLS_axis},  we have positioned the observer at $r_{obs}=20\,M$, $\theta_{\textrm{obs}}=5^\circ$.}
\label{Lensing_EFLS_thin_axis}
\end{figure*}

\subsection{Astrophysical Images of E-FLS stars surrounded by an optically thick disk}
Let us first consider a geometrically thin and opaque accretion disk (also known as optically thick) surrounding the E-FLS star. The accretion disk is located along the equatorial plane. Due to the opacity, we evolve the light rays from the observer's position until they escape to spatial infinity or intersect the accretion disk. Moreover, we consider that the disk is constituted by massive particles following circular geodesic motion, as described in Sec~\ref{Subsec_timelike}. The energy, angular momentum and orbital frequency for a particle at radius $r_c$ are given by Eqs.~\eqref{E-L-timelike} and \eqref{Omega-timelike}, respectively. 

For a geometrically thin and optically thick accretion disk, the solution for the radiative transfer equation~\eqref{RTE} is quite straightforward. We consider that the accretion disk's emission is monochromatic with frequency $\nu_{em}$:
\begin{align}
\label{em_profile}I_\nu\propto \delta(\nu-\nu_{em})\,\epsilon(r),
\end{align}
as measured by an observer comoving with the accretion disk. Moreover based on Refs.~\cite{Proca_Star_Shadow, Ivo_Proca_Shadow}, we assume that $\epsilon(r)$ behaves as follows:
\begin{align}
\label{em_profile2}\epsilon(r)\equiv \frac{1+\tanh[50(r-6M)]}{2}\left(\frac{6\,M}{r}\right)^3.
\end{align}
This expression serves as an ad-hoc emission profile, yet effectively captures the essential features of more intricate emission profiles.
The specific intensity as measured by the observer can be obtained from the invariant intensity~\eqref{invar_RTE}, namely
\begin{align}
\label{Iobs-Iem}I^{obs}_{\nu'}=\frac{\nu'^3}{\nu^3}I_{\nu},
\end{align}
where $\nu'$ is the observed frequency. 

In Fig.~\ref{Lensing_EFLS}, we show the astrophysical images for the different E-FLS configurations (presented in Table~\ref{Sols}) surrounded by a geometrically thin and optically thick accretion disk. In Fig.~\ref{Lensing_EFLS}, we considered the observer at a radius $r_{obs}=20\,M$ and the polar angle $\theta_{\textrm{obs}}=80^\circ$, hence the observer is slightly displaced from the equatorial plane. We notice that the gravitational lensing effect is not strong for configurations FLS1 and FLS3, as can be seen in panels (a) and (c) of Fig.~\ref{Lensing_EFLS}, where the image of the accretion disk is slightly distorted. These results are in accordance with the fact that configurations FLS1 and FLS3 are not compact enough to possess light rings. On the other hand, we notice a strong gravitational lensing effects for the configurations FLS2 and FLS4, as shown in panels (b) and (d) of Fig.~\ref{Lensing_EFLS}. In particular, the gravitational lensing effect in this configuration is so pronounced that we can observe the back part of the accretion disk being gravitationally lensed, appearing both above and below the center of the final image. These results are in accordance with the fact that configurations FLS2 and FLS4 are more compact and they even allow the presence of a light ring.

Furthermore, as depicted in Fig.~\ref{Lensing_EFLS}, we note that the right side of the image appears brighter than the left side. This discrepancy arises from the rotation of the accretion disk, inducing a Doppler effect for the observer. The Doppler effect and the invariance of $I_\nu/\nu^3$ cause the difference in the observed brightness on each side of the accretion disk~\cite{Misner_book}.

We also obtained the astrophysical images of E-FLS stars as viewed by an observer positioned at an observational angle $\theta_{obs}=5^\circ$, which corresponds to a face-on perspective of the disk. The motivation for this choice of angle is based on the recent EHT results that disfavor observations at a high observational angles~\cite{M87_1:2019, M87_2:2019, M87_3:2019, M87_4:2019, M87_5:2019, M87_6:2019, sgra_1:2022, sgra_2:2022, sgra_3:2022, sgra_4:2022, sgra_5:2022, sgra_6:2022}. For instance, in the case of Sgr A*, observational angles $\theta_{obs}>50^\circ$ are disfavored by the comparison between the theoretical model and the observations~\cite{sgra_1:2022}. In Fig.~\ref{Lensing_EFLS_axis}, we show the gravitational lensing for E-FLS stars, as seen by an observer at $r_{obs}=20\,M$ and $\theta_{obs}=5^\circ$. We notice that the images for the different configurations are more similar, in comparison to the images for an observer at the equatorial plane. The major difference relies in the brightness difference for each configuration. The less compact configurations (FLS1 and FLS3) appear brighter than the more compact ones (FLS2 and FLS4). These results arises from the stronger gravitational redshift associated with the more compact configuration, in conjunction with the formula for the invariant intensity, as provided in Eq.~\eqref{Iobs-Iem}.

\subsection{Astrophysical Images of E-FLS stars surrounded by an optically thin disk}
Let us now consider an optically thin accretion disk surrounding the E-FLS star. An optically thin disk is transparent to radiation and the light rays can cross it several times before being scattered. Each time that a given light ray intersects the accretion disk, it acquires more intensity. We consider the same emission profile as given in Eqs.~\eqref{em_profile}-\eqref{em_profile2}. We also keep the position of the observer at $r_{obs}=20\,M$ and vary the polar angle of observation. 

In Fig.~\ref{Lensing_EFLS_thin}, we show the gravitational lensing for the different E-FLS configurations in Table~\ref{Sols}, for a fixed polar observation angle $\theta_{obs}=80^\circ$. We notice that for less compact configurations (FLS1 and FLS3), the images for the thin disk are akin to the images with optically thick disk. However, for more compact configurations that have a light ring (FLS2 and FLS4), the images can be quite different when compared to the optically thick case. The major difference arises due to the thin signatures on the images, dubbed as photon ring~\cite{Universal_Int}. These thin signatures are associated to the existence of light rings. The light rays that have an impact parameter close to the critical one [see Eq.~\eqref{LR_Conds}] complete several turns around the center before being scattered, intersecting the disk many times and consequently increasing the intensity.

In Fig.~\ref{Lensing_EFLS_thin_axis}, we present the gravitational lensing for the E-FLS stars in Table~\ref{Sols} for an observer at $r_{obs}=20\,M$ and $\theta_{obs}=5^\circ$, i.e. an almost face-on observation. We notice that the configurations FLS2 and FLS4 present a photon ring close to the center of the image, while the configurations FLS1 and FLS3 do not. As discussed previously, such thin signature close to the center of the image arises due to the light rays that cross the accretion disk several times for certain values of the impact parameter. 

We notice that the presence of thin photon rings on the image leave a characteristic fingerprint on the interferometric signature on very long baselines~\cite{Universal_Int}. The presence (or the absence) of such photon rings can, in principle, be used as a test for distinguishing different configurations with the future advent of ng-EHT.

\section{Conclusions}
\label{Conclusions}
We have studied the boson star solutions in the context of the E-FLS theory, which we dubbed as E-FLS stars. The matter sector of the Lagrangian in the E-FLS theory is constituted by two interacting scalar fields, one being real and the other complex. By assuming a spherically symmetric ansatz for the background metric and the scalar fields, we computed numerical solutions for the E-FLS field equations. Within our numerical scheme, we were able to obtain a full scan of the solution space in the E-FLS theory.
We explored the E-FLS star solutions ranging from the limit in which $\mu$ approaches infinity, where we recovered mini-boson star solutions, to the $\mu=0$ limit, where the effects of the FLS coupling became pronounced. In particular, we found interesting results for the case $\mu=0$, in which the real scalar field becomes massless.

We computed some properties of the E-FLS stars, such as, the ADM mass, the total Noether charge, the effective radius and the compactness. An intriguing feature of E-FLS stars is that as we decrease the value of $\mu$, the range of frequencies for which E-FLS stars exist increases. Moreover, we also computed the binding energy and showed that the case $\mu=0$ exhibits the largest region with bound configurations and, consequently, the largest region with linearly stable configurations. Concerning the compactness of the E-FLS stars, we showed that the configurations with $\mu=0$ are less compact, in comparison to the cases with $\mu>0$. 

Motivated by the recent EHT image results, we studied the geodesics around the E-FLS stars and the astrophysical images of such compact objects, when surrounded by an accretion disk. We analyzed the timelike geodesics, describing the motion of massive particles around the E-FLS stars. For instance, we computed the angular velocity for timelike circular geodesics. Our findings revealed that more compact configurations exhibit a maximum value for the angular velocity, whereas less compact configurations do not exhibit such a maximum. Concerning the null geodesic motion, we computed the light rings in E-FLS star spacetimes. We obtained that the E-FLS stars close to the newtonian regime do not have light rings. However, as the configurations become more compact, pairs of stable and unstable light rings emerge. This result is a consequence of the conservation of the total topological charge for horizonless exotic compact objects.

We computed the astrophysical images of E-FLS stars using the backwards ray-tracing technique. We assumed that the E-FLS stars were surrounded by two different models of disk, namely, an optically thick (opaque) and an optically thin (transparent). We found that the configuration close to the Newtonian regime do not exhibit a prominent gravitational lensing, neither a prominent gravitational redshift, since the images are brighter. On the other hand, the configurations that are more compact, and posses a light ring, exhibit strong gravitational lensing and gravitational redshift effects in the astrophysical images. For the optically thin accretion model, we noticed that a thin signature is present in the images due to the light rays that orbits several times around the center before being scattered. The images of such compact configuration can, in principle, mimic the visual appearance of a Schwarzschild black hole. 

\begin{acknowledgments}
The authors would like to acknowledge Funda\c{c}\~ao Amaz\^onia de Amparo a Estudos e Pesquisas (FAPESPA), Conselho Nacional de Desenvolvimento Cient\'ifico e Tecnol\'ogico (CNPq) and Coordena\c{c}\~ao de Aperfei\c{c}oamento de Pessoal de N\'ivel Superior (CAPES) -- Finance Code 001, from Brazil, for partial financial support. H.~L.~J. thanks Etevaldo Costa and Pedro Cunha for useful discussions. This work is supported by the Center for Research and Development in Mathematics and Applications (CIDMA) through the Portuguese Foundation for Science and Technology Funda\c c\~ao para a Ci\^encia e a Tecnologia), UIDB/04106/2020, UIDP/04106/2020,
https://doi.org/10.54499/UIDB/04106/2020 
and https://doi.org/10.54499/UIDP/04106/2020.
The authors acknowledge support from the projects
http://doi.org/10.54499/PTDC/FISAST/3041/2020,
http://doi.org/10.54499/CERN/FIS-PAR/0024/2021 and https://doi.org/10.54499/2022.04560.PTDC.  This work has further been supported by the European Horizon Europe staff exchange (SE) programme HORIZON-MSCA2021-SE-01 Grant No. NewFunFiCO101086251.
\end{acknowledgments}

\label{ApxA}

{}

\begin{thebibliography}{99}







\bibitem{LIGO-VIRGO2016} B. P. Abbott et al., Observation of Gravitational Waves from a Binary Black Hole Merger.\href{https://link.aps.org/doi/10.1103/PhysRevLett.116.061102
}{ Phys. Rev. Lett., 116, 061102, (2016).}

\bibitem{LIGO-VIRGO} B. P. Abbott \textit{et al}, GWTC-1: A Gravitational-Wave Transient Catalog of Compact Binary Mergers Observed by LIGO and Virgo during the First and Second Observing Runs. \href{https://link.aps.org/doi/10.1103/PhysRevX.9.031040
}{Phys. Rev. X, \textbf{9}, 031040 (2019).}

\bibitem{M87_1:2019} The Event Horizon Telescope Collaboration, 
		First M87 event horizon telescope results. I. The shadow of the supermassive black hole, 
		\href{https://doi.org/10.3847/2041-8213/ab0ec7}{Astrophys. J. Lett. {\bf 875}, L1 (2019).} 

\bibitem{M87_2:2019} The Event Horizon Telescope Collaboration, 
		First M87 Event Horizon Telescope Results. II. Array and Instrumentation,
		\href{https://doi.org/10.3847/2041-8213/ab0c96}{Astrophys. J. Lett. {\bf 875}, L2 (2019).}

\bibitem{M87_3:2019} The Event Horizon Telescope Collaboration, 
		First M87 Event Horizon Telescope Results. III. Data Processing and Calibration, 
		\href{https://doi.org/10.3847/2041-8213/ab0c57}{Astrophys. J. Lett. {\bf 875}, L3 (2019).}

\bibitem{M87_4:2019} The Event Horizon Telescope Collaboration, 
		First M87 Event Horizon Telescope Results. IV. Imaging the Central Supermassive Black Hole, 
		\href{https://doi.org/10.3847/2041-8213/ab0e85}{Astrophys. J. Lett. {\bf 875}, L4 (2019).}

\bibitem{M87_5:2019} The Event Horizon Telescope Collaboration, 
		First M87 Event Horizon Telescope Results. V. Physical Origin of the Asymmetric Ring, 
		\href{https://doi.org/10.3847/2041-8213/ab0f43}{Astrophys. J. Lett. {\bf 875}, L5 (2019).}

\bibitem{M87_6:2019} The Event Horizon Telescope Collaboration, 
		First M87 Event Horizon Telescope Results. VI. The Shadow and Mass of the Central Black Hole, 
		\href{https://doi.org/10.3847/2041-8213/ab1141}{Astrophys. J. Lett. {\bf 875}, L6 (2019).}
		
		
\bibitem{sgra_1:2022} The Event Horizon Telescope Collaboration, 
		First Sagittarius A* Event Horizon Telescope Results. I. The Shadow of the Supermassive Black Hole in the Center of the Milky Way,
		\href{https://doi.org/10.3847/2041-8213/ac6674}{Astrophys. J. Lett. {\bf 930}, L12 (2022).}
		
\bibitem{sgra_2:2022} The Event Horizon Telescope Collaboration, 
		First Sagittarius A* Event Horizon Telescope Results. II. EHT and Multiwavelength Observations, Data Processing, and Calibration, 
		\href{https://doi.org/10.3847/2041-8213/ac6675}{Astrophys. J. Lett. {\bf 930}, L13 (2022).}
		
\bibitem{sgra_3:2022} The Event Horizon Telescope Collaboration, 
		First Sagittarius A* Event Horizon Telescope Results. III. Imaging of the Galactic Center Supermassive Black Hole, 
		\href{https://doi.org/10.3847/2041-8213/ac6429}{Astrophys. J. Lett. {\bf 930}, L14 (2022).}

\bibitem{sgra_4:2022} The Event Horizon Telescope Collaboration,
		 First Sagittarius A* Event Horizon Telescope Results. IV. Variability, Morphology, and Black Hole Mass, 
		 \href{https://doi.org/10.3847/2041-8213/ac6736}{Astrophys. J. Lett. {\bf 930}, L15 (2022).}
		
\bibitem{sgra_5:2022} The Event Horizon Telescope Collaboration,
		 First Sagittarius A* Event Horizon Telescope Results. V. Testing Astrophysical Models of the Galactic Center Black Hole,
		  \href{https://doi.org/10.3847/2041-8213/ac6672}{Astrophys. J. Lett. {\bf 930}, L16 (2022).}
		  
		
\bibitem{sgra_6:2022} The Event Horizon Telescope Collaboration, 
		First Sagittarius A* Event Horizon Telescope Results. VI. Testing the Black Hole Metric, 
		\href{https://doi.org/10.3847/2041-8213/ac6756}{Astrophys. J. Lett. {\bf 930}, L17 (2022).}


\bibitem{Kerr_Hypothesis} C. A. R. Herdeiro, Black Holes: On the Universality of the Kerr Hypothesis, \href{https://doi.org/10.1007/978-3-031-31520-6_8}{Lect. Notes Phys. \textbf{1017}, 315 (2023).}

\bibitem{LeePang} T. D. Lee and P. Yang, Stability of mini-boson stars. \href{https://doi.org/10.1016/0550-3213(89)90365-9}{Nucl. Phys. B \textbf{315} (1989).}

\bibitem{Santos:2024vdm}
N.~M.~Santos, C.~L.~Benone and C.~A.~R.~Herdeiro,
``Radial stability of spherical bosonic stars and critical points,''
[arXiv:2404.07257 [gr-qc]].

\bibitem{grav_cool1} Edward Seidel and Wai-Mo Suen. Formation of solitonic stars through gravitational cooling.
\href{https://doi.org/10.1103/PhysRevLett.72.2516}{Phys. Rev. Lett., \textbf{72} 2516, (1994).}

\bibitem{grav_cool2} F. Siddhartha Guzman and L. Arturo Urena-Lopez. Gravitational cooling of self-gravitating Bose-Condensates. \href{https://dx.doi.org/10.1086/504508}{Astrophys. J., \textbf{645}, 814, (2006).}

\bibitem{grav_cool3} F. Giovanni, N. S. G., C. A. R. Herdeiro, and J. A. Font. Dynamical formation of Proca stars and quasistationary solitonic objects. \href{https://doi.org/10.1103/PhysRevD.98.064044}{Phys. Rev. D, \textbf{98} 064044, (2018).}

\bibitem{Kaup} D. J. Kaup, Klein-gordon geon. \href{https://doi.org/10.1103/PhysRev.172.1331}{Phys. Rev. \textbf{172}, 1331 (1968).}


\bibitem{Proca1}R. Brito, V. Cardoso, C. A. R. Herdeiro, and E. Radu. Proca stars: Gravitating Bose–Einstein condensates of massive spin 1 particles. \href{https://doi.org/10.1016/j.physletb.2015.11.051}{Phys. Lett. B, \textbf{752}, 291, 2016.}

\bibitem{Proca2}C. Herdeiro, I. Perapechka, E. Radu, and Ya. Shnir. Asymptotically flat spinning scalar, Dirac and Proca stars. \href{https://doi.org/10.1016/j.physletb.2019.134845}{Phys. Lett. B, \textbf{797}, 134845, 2019.}

\bibitem{FLS} R. Friedberg, T. D. Lee and A. Sirlin, Class of scalar-field soliton solutions in three space dimensions. \href{https://doi.org/10.1103/PhysRevD.13.2739}{Phys. Rev. D \textbf{13}, 2739 (1976).}

\bibitem{Rotating_FLS}J. Kunz, I. Perapechka, and Y. Shnir, Kerr black holes with synchronised scalar hair and boson stars in the Einstein-Friedberg-Lee-Sirlin model, \href{https://doi.org/10.1007/JHEP07(2019)109}{JHEP. \textbf{07}, 109, (2019).}

\bibitem{U1_FLS} J. Kunz, V. Loiko, and Ya. Shnir, $U(1)$ gauged boson stars in the Einstein-Friedberg-Lee-Sirlin model. \href{https://doi.org/10.1103/PhysRevD.105.085013}{Phys. Rev. D \textbf{105}, 085013, 2022.}

\bibitem{Proca-Higgs} C. Herdeiro, E. Radu, and E. dos Santos Costa Filho, Proca-Higgs balls and stars in a UV completion for Proca self-interactions, \href{https://dx.doi.org/10.1088/1475-7516/2023/05/022}{JCAP \textbf{05}, 022 (2023).}

\bibitem{colsys} U. Ascher, J. Christiansen, and R. D. Russell, A collocation solver for mixed order systems of boundary value problems.\href{https://doi.org/10.1090/S0025-5718-1979-0521281-7}{Math. Comput. \textbf{33}, 659 (1979).}

\bibitem{Cpp} W. H. Press, S. A. Teukolsky, W.T. Vetterling. \textit{Numerical recipes in C++: The art of scientific computing} (Cambridge University Press, 2002).

\bibitem{Cunha2017} P. V. P. Cunha, E. Berti and C. A. R. Herdeiro, Light-Ring Stability for Ultracompact Objects.\href{https://doi.org/10.1103/PhysRevLett.119.251102}{Phys. Rev. Lett. \textbf{119}, 251102 (2017).}

\bibitem{Rubakov2011} A. Levin and V. Rubakov, Q-Balls with scalar charge.\href{https://doi.org/10.1142/S0217732311034992}{Mod. Phys. Lett. A \textbf{26}, 409 (2011).}

\bibitem{Loiko2018} V. Loiko, I. Parapechka and Ya. Shnir, Q
-balls without a potential.\href{https://doi.org/10.1103/PhysRevD.98.045018}{Phys. Rev. D \textbf{98}, 045018 (2018).}

\bibitem{Cunha2020} P. V. P. Cunha and C. A. R. Herdeiro, Stationary Black Holes and Light Rings.\href{https://doi.org/10.1103/PhysRevLett.124.181101}{Phys. Rev. Lett.~\textbf{124}, 181101 (2020).}

\bibitem{Lindquist} R. W. Lindquist, Relativistic transport theory. \href{https://doi.org/10.1016/0003-4916(66)90207-7}{Annals of Physics, \textbf{37}, 487 (1966).}

\bibitem{Bohn2015} A. Bohn, W. Throwe, F. Hbert, K. Henriksson, and D. Bunandar, What does a binary black hole merger look like? \href{https://doi.org/10.1088/0264-9381/32/6/065002}{ Classical Quant. Grav. \textbf{32}, 065002 (2015).}

\bibitem{Cunha2018} P. V. P. Cunha and C. A. R. Herdeiro, Shadows and strong gravitational lensing: a brief review. \href{https://link.springer.com/article/10.1007/s10714-018-2361-9}{ Gen. Relativ. Gravit. \textbf{50}, 42 (2018).}

\bibitem{Cunha20202} P. V. P. Cunha, C. A. R. Herdeiro, E. Radu, and H. F. Rúnarsson, Shadows of kerr black holes with scalar hair. \href{https://journals.aps.org/prl/abstract/10.1103/PhysRevLett.115.211102}{ Phys. Rev. Lett. \textbf{124}, 181101 (2020).}

\bibitem{HCDL2021} H. C. D. Lima Junior, P. V. P. Cunha, C. A. R. Herdeiro, and L. C. B. Crispino, Shadows and lensing of black holes immersed in strong magnetic fields.\href{https://journals.aps.org/prd/abstract/10.1103/PhysRevD.104.044018}{ Phys. Rev. D \textbf{104}, 044018 (2021).}

\bibitem{Proca_Star_Shadow} J.~L.~Rosa and D. R. Garcia, Shadows of boson and Proca stars with thin accretion disks. \href{https://journals.aps.org/prd/abstract/10.1103/PhysRevD.106.084004}{ Phys. Rev. D \textbf{106}, 084004 (2022).}

\bibitem{Ivo_Proca_Shadow} I. Sengo, P. V. P. Cunha, C. A. R. Herdeiro and E. Radu, The imitation game reloaded: effective shadows of dynamically robust spinning Proca stars. \href{https://arxiv.org/abs/2402.14919}{arXiv:2402.14919.}

\bibitem{Caio_shadows} J. L. Rosa, C. F. B. Macedo and D. R. Garcia, Imaging compact boson stars with hot spots and thin accretion disk. \href{https://doi.org/10.1103/PhysRevD.108.044021}{Phys. Rev. D \textbf{108}, 044021 (2023)}.

\bibitem{Misner_book} C. W. Misner, K. S. Thorne e J. A. Wheeler, \href{}{Gravitation (Freeman, San Francisco, 1973).}

\bibitem{Universal_Int} M.~D.~Johnson, A.~Lupsasca, A.~Strominger et al, Universal interferometric signatures of a black hole’s photon rin. \href{https://www.science.org/doi/10.1126/sciadv.aaz1310}{
Science Advances \textbf{6}, eaaz1310  (2020).}

\bibitem{Coleman1986} S. R. Coleman, Q-balls. \href{https://doi.org/10.1016/0550-3213(86)90520-1}{Nucl. Phys. \textbf{B262}, 263 (1985); \textbf{269}, 744(E) (1986).}
\end{thebibliography}
\end{document}